\documentclass[letterpaper,superscriptaddress,english,aps,prx,twocolumn, amsfonts, amssymb]{revtex4}
\usepackage{epsfig, graphicx,graphics,amsmath,amssymb,float}
\usepackage[T1]{fontenc}
\usepackage[latin9]{inputenc}
\usepackage{amsmath}
\usepackage{amssymb}

\usepackage{amscd}
\usepackage{bm}
\usepackage{psfrag}

\usepackage{babel}

\usepackage{bbm} 

\usepackage{color}
\usepackage[bookmarks=true,colorlinks,linkcolor=blue,urlcolor=blue,citecolor=blue]{hyperref}

\newcommand{\be}{\begin{equation}}
\newcommand{\ee}{\end{equation}}
\newcommand{\bea}{\begin{eqnarray}}
\newcommand{\eea}{\end{eqnarray}}
\newcommand{\ket}[1]{{|#1\rangle}}
\newcommand{\bra}[1]{\langle#1|}
\newcommand{\mc}{\mathcal}

\newcommand{\edit}[1]{{#1}}

\begin{document}

\title{Dynamical structure factors  in the  nematic   phase  of frustrated ferromagnetic spin chains}
\author{Fl\'avia B. Ramos}
\affiliation{
International Institute of Physics, Universidade Federal do Rio Grande do Norte, Campus Universit\'{a}rio, Lagoa Nova, Natal-RN 59078-970, Brazil}
\author{Sebas Eli\"ens}
\affiliation{
International Institute of Physics, Universidade Federal do Rio Grande do Norte, Campus Universit\'{a}rio, Lagoa Nova, Natal-RN 59078-970, Brazil}
\author{Rodrigo G. Pereira}
\affiliation{
International Institute of Physics, Universidade Federal do Rio Grande do Norte, Campus Universit\'{a}rio, Lagoa Nova, Natal-RN 59078-970, Brazil}
\affiliation{Departamento de F\'isica Te\'orica e Experimental, Universidade Federal do Rio Grande do
Norte, 59078-970, Natal, RN, Brazil}

\begin{abstract}
Frustrated spin systems
can show phases with spontaneous breaking of spin-rotational
symmetry without the formation of local magnetic order.  
We  study the dynamic response of the spin-nematic phase of one-dimensional spin-$1/2$ systems, characterized by slow  large-distance decay of quadrupolar correlations, 
by numerically computing   one-spin and two-spin dynamical  structure factors at zero temperature using    
 time-dependent density matrix renormalization group methods.   We interpret the results in terms of an effective theory of
gapped magnon excitations interacting with a quasi-condensate of bound magnon pairs. This employs an extension
of the well-known Tomonaga-Luttinger liquid theory  which includes the magnon states as a mobile impurity. 
 A good qualitative understanding of
the characteristic thresholds and their intensity in the   structure factors   is obtained this way. Our results are useful in the interpretation of inelastic neutron  scattering and resonant inelastic x-ray scattering   experiments.

\end{abstract}
\maketitle

\section{Introduction}

Most open problems in quantum magnetism relate to the search for phases of interacting spin systems that depart from the paradigm of long-range magnetic order \cite{Penc2011}. One example is the spin nematic phase, characterized by a quadrupolar order parameter \cite{Blume1969,Andreev1984}. For spin-1/2 systems, a bond spin-nematic order parameter is defined from  the traceless symmetric rank-2 tensor $Q^{ab}_{ij}=S_{i}^aS_{j}^b+S_i^bS_{j}^a-\frac23\delta^{ab}\mathbf S_i\cdot \mathbf S_{j}$, where $a,b\in\{x,y,z\}$ and $i,j$ label nearest-neighbor spins. In contrast with dipolar magnetic order, where $\langle \mathbf S_i\rangle\neq0$ in the ground state, a nonzero expectation value of any component of $Q^{ab}_{ij}$ breaks spin-rotation invariance but preserves time-reversal symmetry. 

Theoretically, it is well established that a one-dimensional (1D) version of the bond spin-nematic phase exists in the frustrated ferromagnetic spin-1/2 chain in a magnetic field \cite{Chubukov1991,Vekua2007,Kuzian2007,Kecke2007,Hikihara2008,Sudan2009,Sato2009,Nishimoto2012,Ren2012,Sato2013,Starykh2014,Onishi2015,Furuya2017,Parvej2017}. This model describes a quantum spin chain with ferromagnetic nearest-neighbor exchange coupling $J_1<0$ and antiferromagnetic next-nearest-neighbor exchange $J_2>0$. Since in one dimension the continuous spin-rotational symmetry cannot be spontaneously broken, the 1D spin-nematic phase has to be understood in terms of quasi-long-range order of quadrupolar correlations, which decay algebraically with distance and more slowly than dipolar correlations. The quadrupolar nematic phase appears in the parameter regime  $\alpha=J_1/J_2\gtrsim -2.72$ at sufficiently high magnetic field  \cite{Kuzian2007,Kecke2007,Hikihara2008,Sudan2009}. Decreasing the field leads to a  crossover to a spin-density wave (SDW) regime, in which the staggered part of the longitudinal spin correlation function decays more slowly than the quadrupolar correlation. At even lower fields, there is a transition to a vector-chiral phase that breaks bond-parity symmetry \cite{Hikihara2008}. In addition, higher-order multipolar phases exist in the range $-4<\alpha\lesssim -2.72$ \cite{Hikihara2008,Sudan2009}.

Both the spin nematic and SDW regimes are described in the low-energy limit as a Tomonaga-Luttinger (TL) liquid with one gapless bosonic mode and gapped single-spin-flip excitations \cite{Hikihara2008,Sato2009,Furuya2017}. The effective low-energy theory can be derived using  bosonization  in the limit of two weakly coupled Heisenberg chains $|\alpha|\ll 1$. For $|\alpha|$ of order 1, one can alternatively consider the limit of large magnetic fields, just below the saturation field, and regard the TL liquid  in the nematic regime as a quasi-condensate of bound magnon pairs treated as hard-core bosons \cite{Chubukov1991,Hikihara2008,Zhitomirsky2010}.

The frustrated ferromagnetic spin chain model finds a nearly ideal realization in the quasi-1D material LiCuVO$_4$ \cite{Prokofiev2004,Enderle2005,Enderle2010,Mourigal2012,Buttgen2014,Orlova2017}. The estimates for the value of $\alpha$ in this material range from $\alpha\approx -0.4$ \cite{Enderle2005} to $\alpha\approx -2$ \cite{Ren2012}.  Remarkably, a recent nuclear magnetic resonance experiment \cite{Orlova2017} provided compelling evidence for the spin-nematic phase  in a narrow window of magnetic field between the SDW phase and the fully polarized state.

The purpose of this work is to  analyze the   dynamics of  the frustrated ferromagnetic spin chain.   It is well known that dynamical structure factors (DSFs) provide direct information about the  excitation spectrum. For instance,  they  can demonstrate the existence of fractional elementary  excitations, such as spinons in the antiferromagnetic Heisenberg chain \cite{Mourigal2013}.  Spinons have been observed by inelastic neutron scattering experiments in LiCuVO$_4$ at zero magnetic field \cite{Enderle2010}.  The  dynamical spin structure factor in this case has been calculated numerically using a time-dependent density matrix renormalization group (tDMRG) algorithm \cite{Ren2012}. In the nematic phase, however, the spectrum of the frustrated chain is organized in terms of gapped magnons and gapless bound magnon pairs. The low-energy features of the  spin DSF in the   nematic phase have been studied within the TL liquid theory \cite{Sato2009}. The The finite-energy  spectrum was investigated  using \edit{the dynamical density-matrix renormalization group (DDMRG) method} in Ref. \cite{Onishi2015}, but the full intensity plots in the energy-momentum plane were restricted to a low magnetization in the SDW regime. 

Here we use  state-of-the-art tDMRG methods \cite{FeiguintDMRG} to calculate the DSFs of various one-spin and two-spin operators inside the nematic phase.  While one usually focuses on the dynamics of one-spin operators due to \edit{their} relevance for inelastic neutron scattering, two-spin excitations can also be probed by the fast-developing techniques of resonant inelastic x-ray scattering (RIXS) \cite{Haverkort2010,Ament2011}. We interpret our numerical results  in light of the current understanding of dynamical correlations of critical 1D systems beyond the low-energy regime described by TL  liquid theory \cite{Imambekov2012,Pereira2012}.  Our high-resolution results clearly show a small single-magnon gap directly in the transverse  spin DSF. 
The corresponding momentum  is, however, not at the minimum of the magnon band but at shifted momenta where furthermore clear replicas are observed. This is explained within \edit{a} description in terms of magnons interacting with the condensate of bound states as a direct consequence of the effective hard-core repulsion between single magnons and the bound magnon pairs.  Based on an effective impurity model, we compute the theoretical threshold exponents that allow us to understand the qualitative features observed in the tDMRG data. Furthermore, we compute DSFs associated with flipping two spins on neighboring sites.
In contrast to the single magnon excitations, the spectrum of two-spin-flip operators is gapless. From the effective description, this is natural as these operators  probe the creation or annihilation of bound magnon pairs in the condensate. We argue that the computed two-spin structure factors behave qualitatively like one-spin structure factors in XXZ spin chains.

This paper is organized as follows. In Sec. \ref{sec:model}, we define the model and main quantities of interest. In Sec. \ref{sec:spectrum}, we discuss the nature and spectra of excitations as one lowers the field from above the saturation field. This outlines how the physics can be understood in terms of an effective model  of bound magnon pairs and single magnons. In Sec. \ref{sec:mim}, we discuss how the excitation spectrum relates to the thresholds of the different DSFs. For the DSFs probing single magnons, we formulate the effective impurity model that is used to derive the threshold exponents.
In Sec. \ref{sec:numerics}, we present our numerical results along with the interpretation in terms \edit{of} magnons interacting with the bound-state condensate. Finally, we provide concluding remarks in Sec. \ref{sec:conclusion}. \edit{The appendices contain details of the calculation of the bare pair-magnon interaction potential in the effective model and of the threshold exponents.}

\section{Model and dynamical structure factors\label{sec:model}}

The Hamiltonian for the frustrated ferromagnetic spin chain is 
\be
H=\sum_{j=1}^L(J_1\mathbf S_j\cdot \mathbf S_{j+1}+J_2 \mathbf S_j\cdot \mathbf S_{j+2}-h S^z_j),\label{model}
\ee
where $\mathbf S_j$ are spin-1/2 operators, $J_1<0$ and $J_2>0$ are exchange coupling constants, and $h$ is the external magnetic field. \edit{Here we consider periodic boundary conditions $\mathbf S_{j+L}=\mathbf S_{j}$.} This model has a global U(1) symmetry corresponding to the conservation of the total longitudinal magnetization $S^z_{\text{tot}}=\sum_jS^z_j$. 
The ground state phase diagram  as a function of $\alpha=J_1/J_2$ and magnetization  $m=\langle S_j^z\rangle$ can be found in Refs. \cite{Hikihara2008,Sudan2009}. In this work, we set $\alpha=-1$ and consider two values of magnetization: $m=0.4$ in the spin-nematic regime and $m=0.2$ in the SDW regime.  

We are interested in dynamical correlation functions for one-spin operators $S_j^a$ and two-spin operators $S_j^a S_{j+1}^b$, where $a,b\in \{x,y,z\}$. In order to select excitations with well-defined quantum numbers of $S^z_{\text{tot}}$, it is convenient to  choose instead $a,b\in\{+,-,z\}$, with $S_j^\pm=  S_j^x\pm iS^y_{j}$. The DSFs for one-spin operators are defined as \be
S^{\bar a a}( q,\omega)=\int_{-\infty}^{+\infty}dt\,e^{i\omega t}\sum_{r}e^{-iqr}\langle gs|S^{\bar a}_{j+r}(t)S^{  a}_{j}(0)|gs\rangle,\label{onespin}
\ee
where $| {gs}\rangle$ is the exact ground state,  $S^a_{j}(t)=e^{iHt}S^a_{j}e^{-iHt}$ is the   operator evolved in real time, and  we use the notation $\bar{a}=-,+,z$ for $a=+,-,z$, respectively, such that $S^{\bar{a}}_j= (S_j^{a})^\dagger $. 
The expression in Eq. (\ref{onespin}) is   equivalent to the Lehmann representation\be
S^{\bar a{a}}( q,\omega)=\frac{2\pi}{L}\sum_{\alpha}|\langle \alpha | \mc O^{a}_{ q}|{gs}\rangle|^2\delta(\omega-E_\alpha+E_{ {gs}}),
\ee
where $\mc O^a_{q}=\sum_je^{i q j}  S^a_j$   and $|\alpha\rangle$ are exact eigenstates of $H$ with energy $E_\alpha$.  Thus, the support of $S^{\bar a{a}}( q,\omega)$ corresponds to the region of the $(q,\omega)$ plane where there are excitations created by the action of  $S^a_j$ on the ground state that carry  momentum $q$ and energy $\omega$. 

For two-spin operators, we define\bea
\hspace{-0.3cm}S^{\bar a\bar b{a}{b}}( q,\omega)&=&\int_{-\infty}^{+\infty}dt\,e^{i\omega t}\sum_{r}e^{-iqr}\nonumber\\
&&\times\langle gs|S^{\bar a}_{j+r}S^{\bar b}_{j+r+1}(t)S^{ a}_{j}(0)S^{  b}_{j+1}(0)|gs\rangle,\label{twospin}
\eea
which is equivalent to\be
S^{\bar{a}\bar{b}ab}( q,\omega)=\frac{2\pi}{L}\sum_{\alpha}|\langle  \alpha| \mc O^{ab}_{ q}|gs\rangle|^2\delta(\omega-E_\alpha+E_{ {gs}}),
\ee
where $\mc O^{ab}_q=\sum_je^{i q j}  S^a_jS^{b}_{j+1}$. Note that the set of two-spin operators  includes not only the components of the quadrupole moment $Q^{ab}_{j,j+1}$, but also the \edit{antisymmetric} part of the two-spin tensor $\epsilon^{abc}S_j^bS_{j+1}^c$. For the Heisenberg chain ($J_2=0$), the antisymmetric part of the two-spin tensor is related to  the spin current between sites $j$ and $j+1$.  The integrability of the 1D Heisenberg model allows one to compute DSFs  exactly using the algebraic Bethe ansatz, including the case of two-spin operators  \cite{Klauser2011,Klauser2012}.  In Section \ref{sec:numerics}, we will show  results for  DSFs for the nonintegrable model with $J_2\neq0$ in the nematic/SDW phase obtained numerically  using  tDMRG.

\section{General properties of the excitation spectrum\label{sec:spectrum}}

\subsection{Spectrum above the saturation field}
To understand the excitation spectrum of the spin nematic phase beyond the low-energy regime, we start from the limit of large magnetic fields. For a fixed value of $\alpha$, there is a saturation field $h_{\text{sat}}(\alpha)$ such that  the exact ground state for $h>h_{\text{sat}}$ is the fully polarized state, $|\Uparrow\rangle =\bigotimes_j {|\uparrow\rangle}_j$. The excitations in this phase are gapped magnons and magnon bound states. The single-magnon state with momentum $k$ is given by \be
|k\rangle = \frac{1}{\sqrt{L}}\sum_{j=1}^Le^{ikj}S_j^-|\Uparrow\rangle.\label{onemagnon}
\ee
\edit{Periodic boundary conditions quantize the momenta} as $k=2\pi n/L$, with $n=1,\dots,L$. 
The magnon dispersion, \be
\varepsilon(k)=J_1[\cos(k)-1]+J_2[\cos(2k)-1]+h,\label{magnondispersion}
\ee
has minima  at $k_0=\cos^{-1}(|J_1|/4J_2)$ and $\bar{k}_0 = 2\pi -k_0$. The wave number $k_0$ is related to the pitch angle of the helical order in the classical model \cite{Hikihara2008}.  The value of $k_0$ is in general incommensurate, but it approaches $k=\pi/2$ in the limit  $J_2\gg |J_1|$ treated by weak-coupling bosonization \cite{Hikihara2008}. The magnon gap is \be
\Delta=\varepsilon(k_0)=-\frac{J_1^2}{8J_2}-J_2+h.
\ee

\begin{figure}[t]
\begin{center}
\includegraphics[width=0.85\columnwidth]{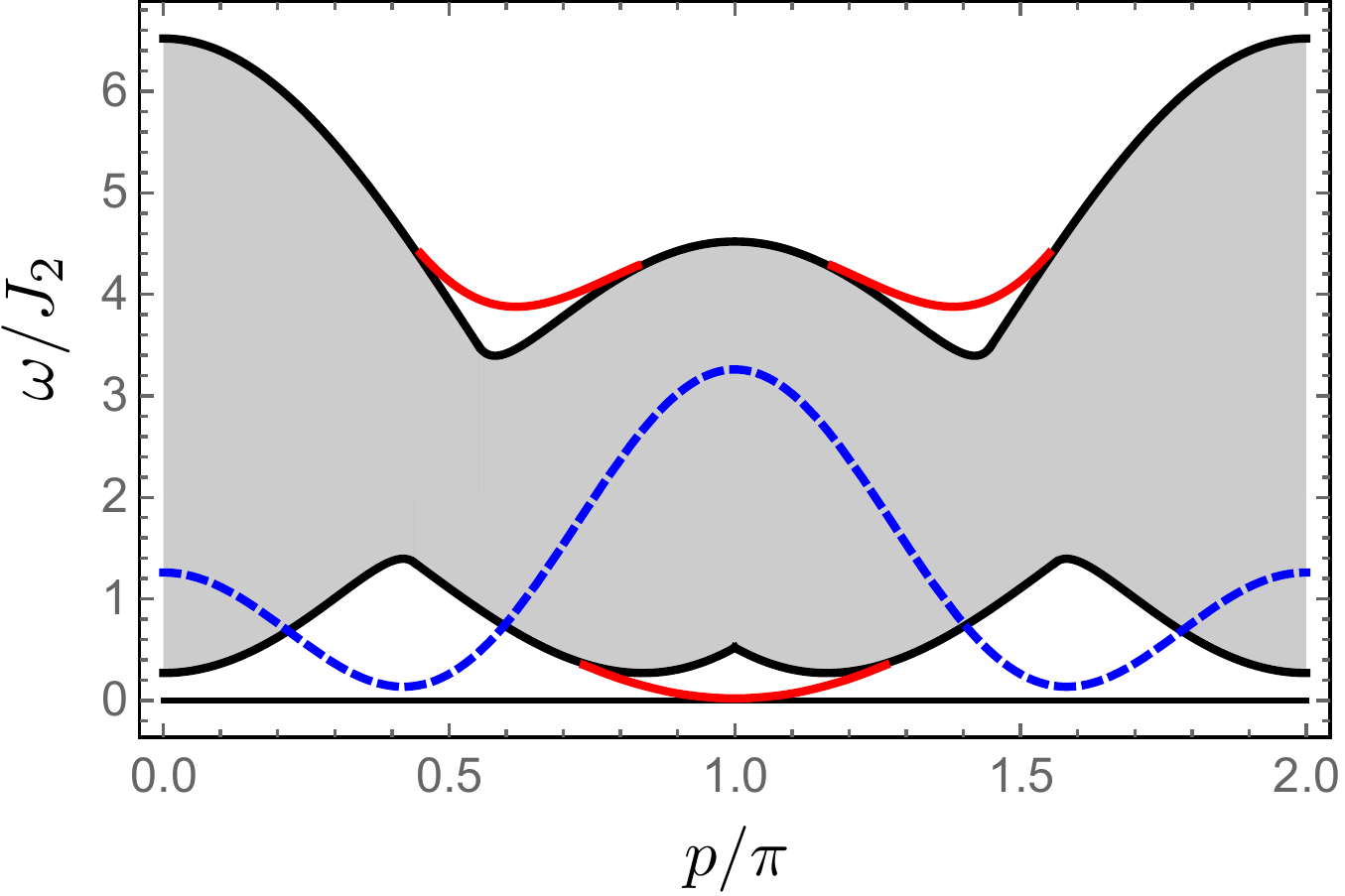}
\end{center}
\caption{One- and two-magnon spectrum   for $\alpha=-1$ and $h=1.26$, slightly  above the saturation field $h_{\text{sat}}\approx 1.25$. The red lines show the dispersion of the  bound magnon pair present both below and above the two-magnon continuum (shaded area).  The dashed blue line represents the single-magnon dispersion.  }
\label{fig:twomagspec}
\end{figure}

The two-particle subspace is spanned by the basis $S^-_{j_1}S^-_{j_2}|\Uparrow\rangle$, $ 1\leq j_1<j_2\leq L$, with  dimension $L(L-1)/2$.  Equivalently, we can use the notation\be
|l,r\rangle =S_l^-S_{l+r}^-|\Uparrow\rangle,
\ee
where $  l=1,\dots,L$ and  $r=1,\dots, (L-1)/2$; here we restrict ourselves to  odd values of $L$ for simplicity. The coordinate $r$ can be interpreted as the relative distance between the two magnons. A two-magnon state with center-of-mass momentum $p$ can be  written as\be
|p,r\rangle =\frac{1}{\sqrt{L}}\sum_{l=1}^{L}e^{ip(l+r/2)}|l,r\rangle,\label{Prbasis}
\ee
where $p=2\pi n/L$ with $n=1,\dots,L$. The matrix elements of the Hamiltonian in the basis of Eq. (\ref{Prbasis}) take the form\be
\langle p',r'|H|p,r\rangle =\delta_{pp'}h_p(r',r),
\ee 
where \be
h_p(r',r)=e^{ip(r-r')/2}\sum_{n=1}^Le^{-ipn}\langle n,r'|H|0,r\rangle.
\ee
The nonzero matrix elements are\begin{align}
&h_p(r,r)=J_1(\delta_{r,1}-2)+J_2(\delta_{r,1}\cos p+\delta_{r,2}-2),\nonumber\\
&h_p(r,r+1)=h_p(r+1,r)=J_1\cos(p/2),\nonumber\\
&h_p(r,r+2)=h_p(r+2,r)= J_2\cos p.
\end{align}

\begin{figure}[t]
\begin{center}
\includegraphics[width=0.85\columnwidth]{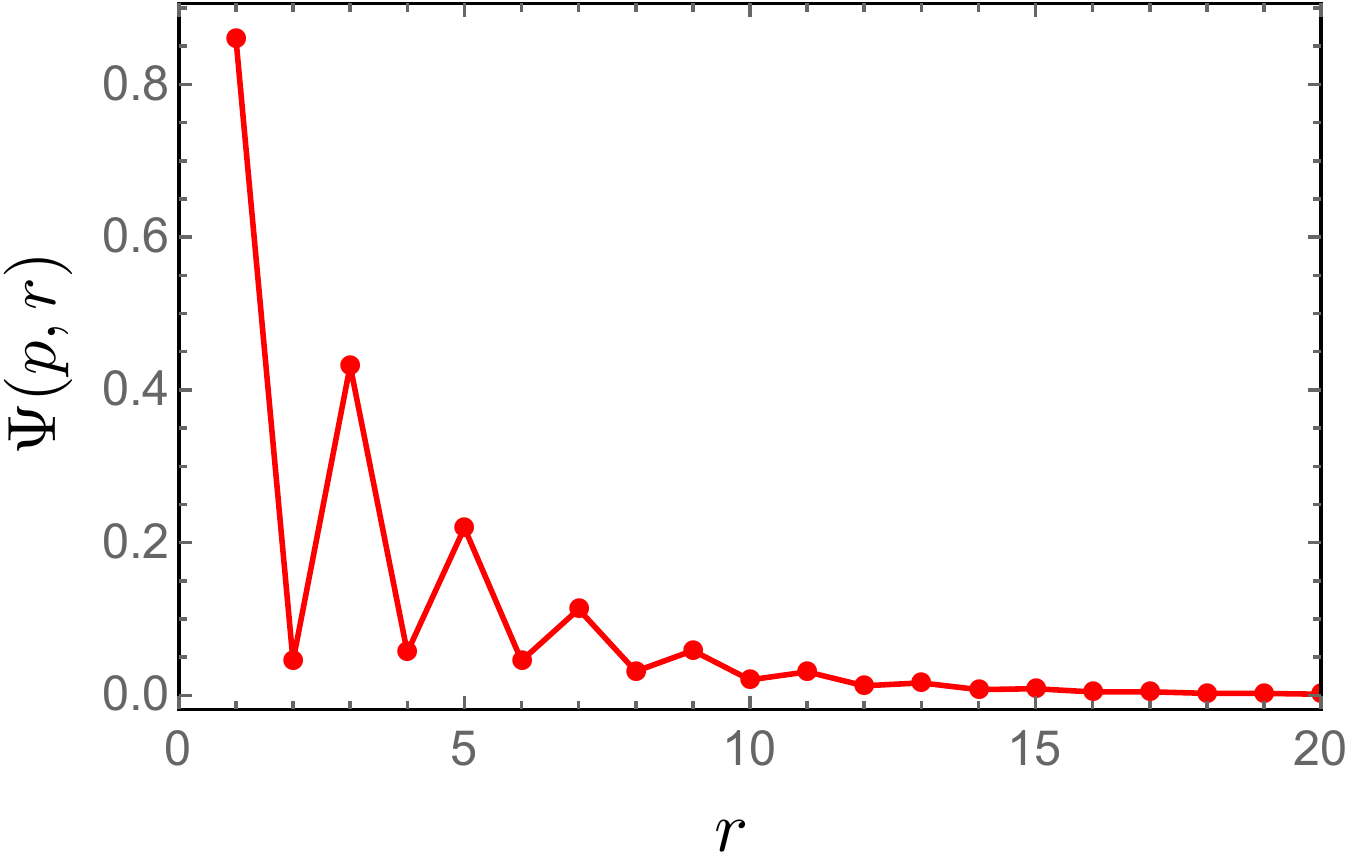}
\end{center}
\caption{Wave function of the two-magnon bound state for $\alpha=-1$ and center-of-mass momentum $p=0.95\pi$.}
\label{fig:wavefunc}
\end{figure}

We find the two-magnon spectrum by  diagonalizing the matrix $h_p(r',r)$ numerically  following Ref. \cite{Kecke2007}. Figure \ref{fig:twomagspec} shows the two-magnon spectrum for  $\alpha=-1$ and $h>h_{\text{sat}}$.  
The main feature is the presence of a  two-magnon bound state both below and above the two-magnon continuum. The bound state dispersion below the continuum, which we denote by $\mc E_b(p)$, has a minimum at $p=\pi$. The bound state dispersion merges with the continuum for $|p-\pi|\geq Q_c$, where $Q_c\approx 0.27\pi$ for $\alpha=-1$. The wave function $\Psi(p,r)$ for the relative coordinate of \edit{the} two-magnon  bound state is illustrated in Fig. \ref{fig:wavefunc}. The corresponding state in the Hilbert space is
\edit
{\be
|b,p\rangle=\sum_{r=1}^{(L-1)/2}\Psi(p,r)|p,r\rangle.\label{boundstates}
\ee
}
Note that the bound state wave function for $p\approx \pi$ has dominant weight at odd values of $r$. Exactly at $p=\pi$, the wave function vanishes at even distances. This     indicates that bound magnon pairs  contribute more to \edit{the}  DSF of odd-distance two-spin operators, such as $S_j^-S_{j+1}^-$, than to even-distance ones like $S_j^-S_{j+2}^-$.

As $h$ approaches the saturation field from above, the minimum energy $\mc E_b(\pi)$ of bound magnon pairs  becomes lower than the minimum energy of a single magnon (see Fig. \ref{fig:twomagspec}). The physical reason is that the ferromagnetic exchange coupling ${J_1<0}$ amounts to an attractive interaction between magnons \cite{Chubukov1991}.  For  $-2.72\lesssim \alpha\lesssim -2.67$, the minimum in the bound state dispersion moves to an incommensurate value of momentum \cite{Kecke2007,Kuzian2007}.  In the range $-4<\alpha\lesssim -2.72$, the interaction becomes strong enough that multi-magnon bound states have even lower energy than the magnon pair.   In this work, we focus on the  regime $-2.67\lesssim\alpha< 0$, in which the gap in the two-magnon bound state dispersion closes at $p=\pi$   as $h\to h_{\text{sat}}^+$.

\subsection{Spectrum below the saturation field}

\subsubsection{Tomonaga-Luttinger liquid  theory and static correlations}
The quadrupolar spin-nematic phase arises for $h<h_{\text{sat}}$ when the gap in $\mc E_b(p)$ closes and  bound magnon pairs condense before single-magnon excitations. Since each pair carries spin eigenvalue $S^z=-2$, the average density of pairs in the ground state, $\rho_0$, is related to the magnetization $m$ by \be
\label{rho0}
\rho_0=\frac12\left(\frac12-m\right).
\ee
In the vicinity of the saturation field from below, we can treat the system as a dilute liquid of bound magnon pairs, $\rho_0\ll1$, with repulsive interactions \cite{Hikihara2008}.

Within a phenomenological harmonic-fluid approach, the   large-distance behavior of   correlation functions of a 
1D Bose liquid  is described  by the TL model \cite{Haldane1981,Cazalilla2011}\be
H_{\text{TL}}=\frac{v}2\int dx\, \left[K(\partial_x\theta)^2+\frac1K(\partial_x\phi)^2\right], \label{TLmodel}
\ee
where $v$ is the sound velocity, $K$ is the Luttinger parameter, $\theta(x)$ is the phase field operator, and $\partial_x\phi$ is associated with  density fluctuations. The bosonic fields obey the commutation relation $[\phi(x),\partial_{x'}\theta(x')]=i\delta(x-x')$. In the   limit $h\to h_{\text{sat}}^-$, we have $K\to1$, the value for hard-core bosons. In the 1D liquid of bound magnon pairs, the correlation functions that have the slowest decay at large distances are \cite{Hikihara2008}\begin{align}
&\langle S^z_{j+r}S^{z}_j\rangle \sim \frac{\cos(2\pi \rho_0 r)}{r^{2K}},\label{Cdip}\\
&\langle S^+_{j+r}S^{+}_{j+r+1}S^-_{j}S^{-}_{j+1}\rangle\sim\frac{(-1)^r}{r^{1/(2K)}}.\label{Cquad}
\end{align}
The spin-nematic regime  is   defined as the region  in the critical phase corresponding to  $K>1/2$, in which the quadrupolar correlation in Eq. (\ref{Cquad}) decays more slowly than the  dipolar correlation in Eq. (\ref{Cdip}). For $K< 1/2$, the system is in the SDW regime in which the longitudinal spin correlation dominates at large distances. On the other hand, the transverse spin correlation decays exponentially:\be
\langle S^+_{j+r}S^{-}_j\rangle \sim e^{-r/\xi}.
\ee
The correlation length $\xi$ is of the order of the inverse gap for single-magnon excitations.

\subsubsection{Beyond the linear dispersion approximation\label{subsec:beyond}}

We now wish to write down an effective model that captures  the support of DSFs beyond the low-energy approximations of the TL model. Our approximations will be justified in the low-density limit $\rho_0\ll1$. In this limit, we assume that the effective Hamiltonian contains only two-particle interactions and we can neglect three-particle scattering processes. Here, two particles can mean two magnons, two bound magnon pairs, or a magnon and a bound magnon pair. Note that the model in Eq. (\ref{model}) has only one U(1) symmetry and only     $S^z_{\text{tot}}$ is a good quantum number. Thus, strictly speaking, the number of magnons and the number of bound magnon pairs are not separately conserved. However, it is known that the vicinity of   thresholds of  spectral functions of critical 1D systems can be described by considering a small, fixed number of elementary excitations at finite energies which  interact with the low-energy modes of the TL liquid \cite{Imambekov2012}. In the following we will apply the same rationale to the spin-nematic phase. 

We start with the effective Hamiltonian for gapped magnons.  
 The scattering of  magnons at low densities is known  to be approximately described by 
an effective  Hamiltonian that includes  one-body and two-body operators in the form  \cite{Feynman1998}
\bea
H_m &=&\sum_{k}\varepsilon(k)a^\dagger_{k}a^{\phantom\dagger}_k\nonumber\\
&&+ \frac{1}{2L}\sum_{k,k',q}V_m(k,k',q)a^\dagger_{k+q}a^{\dagger}_{k'-q}a^{\phantom\dagger}_{k'}a^{\phantom\dagger}_k,\label{effHmagnon}
\eea
where $a_k$ annihilates a magnon with momentum $k$ and energy $\varepsilon(k)$. The effective scattering amplitude  can be extracted from the original spin Hamiltonian by computing  the matrix element  \cite{Feynman1998}
\be
V_m(k,k',q)=L\langle k+q,k'-q|\delta H_{mm}(k,k')|k,k'\rangle, \label{Vmagnon}
\ee
where $|k,k'\rangle$ are two-magnon states [tensor product of states in Eq. (\ref{onemagnon})] and $\delta H_{mm}(k,k')=H-E_0 -\varepsilon (k)-\varepsilon (k')$.  Here $E_0$ is the energy of the fully polarized state, which plays the role of the vacuum. The subtraction of the energy of the two free magnons in $\delta H_{mm}(k,k')$ is equivalent to dropping the disconnected Feynman diagrams in the four-point function; it is necessary because the scattering states $|k,k'\rangle$ and $|k+q,k'-q\rangle$ are not orthogonal for finite size $L$ \cite{Feynman1998}. We recall that  magnons have to be treated as hard-core bosons. The expression in Eq. (\ref{Vmagnon}) accounts for the regular, finite-range part of the interaction potential.   

Let us now  consider the effective Hamiltonian for bound magnon pairs, of which we have a finite but low density in the ground state. The  Hilbert space of a single pair is spanned by the states $|b,p\rangle$ in Eq. (\ref{boundstates}). In analogy with Eq. (\ref{effHmagnon}), we write down the effective  Hamiltonian with one-body and two-body operators
\be
H_b =\sum_{p}\mc E_b(p)b^\dagger_{p}b^{\phantom\dagger}_p+ \frac{1}{2L}\sum_{p,p',q}V_b(p,p',q)b^\dagger_{p+q}b^{\dagger}_{p'-q}b^{\phantom\dagger}_{p'}b^{\phantom\dagger}_p,\label{effHpair}
\ee
 where $b_p$  annihilates a bound magnon pair with momentum $p$ and energy   $\mc E_b(p) $, and $V_b(p,p',q)$ is the effective scattering amplitude
 \be
V_b(p,p',q)=L\langle b,p+q;b,p'-q|\delta H_{bb}(p,p')|b,p;b,p'\rangle, \label{Vpair}
\ee
with $\delta H_{bb}(p,p')=H-E_0 -\mc E_b (p)-\mc E_b (p')$.
 We note that all momenta of bound magnon pairs  must be  restricted to the interval $[\pi-Q_c,\pi+Q_c]$.

For $h>h_{\text{sat}}$, the pair dispersion    in the vicinity of $p=\pi$ can be written as \be
\mc E_b(p\approx \pi) = \mc E_b(\pi)+\frac{(p-\pi)^2}{2M}+\dots,\label{pairmass}
\ee
where \edit{$M=[(d^2\mathcal{E}_b/dp^2)|_{p=\pi}]^{-1}$} is the effective mass of the bound magnon pair. We can interpret $\mu\equiv-\mc E_b(\pi)$ as the chemical potential for the  magnon pairs. For $h<h_{\text{sat}}$, these bosonic excitations condense, and the pair-pair interaction is responsible for changing  the low-energy dispersion from  quadratic to linear. In spatial dimension higher than one, the linear dispersion of a superfluid phase is qualitatively  described by the Bogoliubov approximation. However,   the assumptions of Bogoliubov theory break down in one dimension \cite{Cazalilla2011}. To understand  how the linear dispersion develops in the 1D liquid of bound magnon pairs, let us consider the asymptotic low-density limit $\rho_0\ll 1$ and focus on  single-particle states with momentum $|p-\pi|\ll Q_c$. In this limit, the average distance between two pairs, $d\sim \rho_0^{-1}$, is  large compared to the size of the bound state, $\ell \sim{ [Q_c-|p-\pi|]^{-1}}$. We can then  introduce the  pair field operator in the continuum limit as\be 
b(x)=\frac1{\sqrt L}\sum_{|q|\ll 1}e^{iqx}b_{\pi+q}.\label{bcontinuumpi}\ee 
Furthermore, if we   approximate the interaction potential by  a contact interaction, $V_b(p,p',q)\approx  V_b(\pi,\pi,0)\equiv   U_0$,  the effective Hamiltonian becomes  simply  \bea
 \label{Heffb}
 H_b&\approx &\int_0^L dx\,\left(-\frac1{2M} b^\dagger\partial_x^2b
\right),\label{Hbcontinuum}
\eea
where we dropped $s$-wave scattering amplitude $U_0$ due to  the hard-core constraint $[b(x)]^2=0$. As a consequence, in the low-density limit the 1D liquid of  bound magnon pairs becomes equivalent to  a Tonks-Girardeau gas   \cite{Girardeau1960},  i.e., the Lieb-Liniger model with infinitely strong repulsion  \cite{LiebLiniger1963,Lieb1963}. This model can be mapped to noninteracting spinless fermions by defining the new fields 
\edit{
\be
b_F(x)=b(x)\cos\left[\pi \int_{-\infty}^{x}dx'\,b^\dagger(x')b(x')\right].\label{JWb}
\ee
}
Here we used the cosine function to keep the Jordan-Wigner string manifestly hermitian in the continuum limit. While the field $b_F(x)$ anticommutes with itself at different positions,   the local density is invariant, $b^\dagger(x) b(x)=b^\dagger_F(x)b^{\phantom\dagger}_F(x)$. Since the fermionic wave function vanishes when two particles occupy  the same position, the hard-core constraint is automatically satisfied for $b_F(x)$.

\begin{figure}[t]
\begin{center}
\includegraphics[width=0.85\columnwidth]{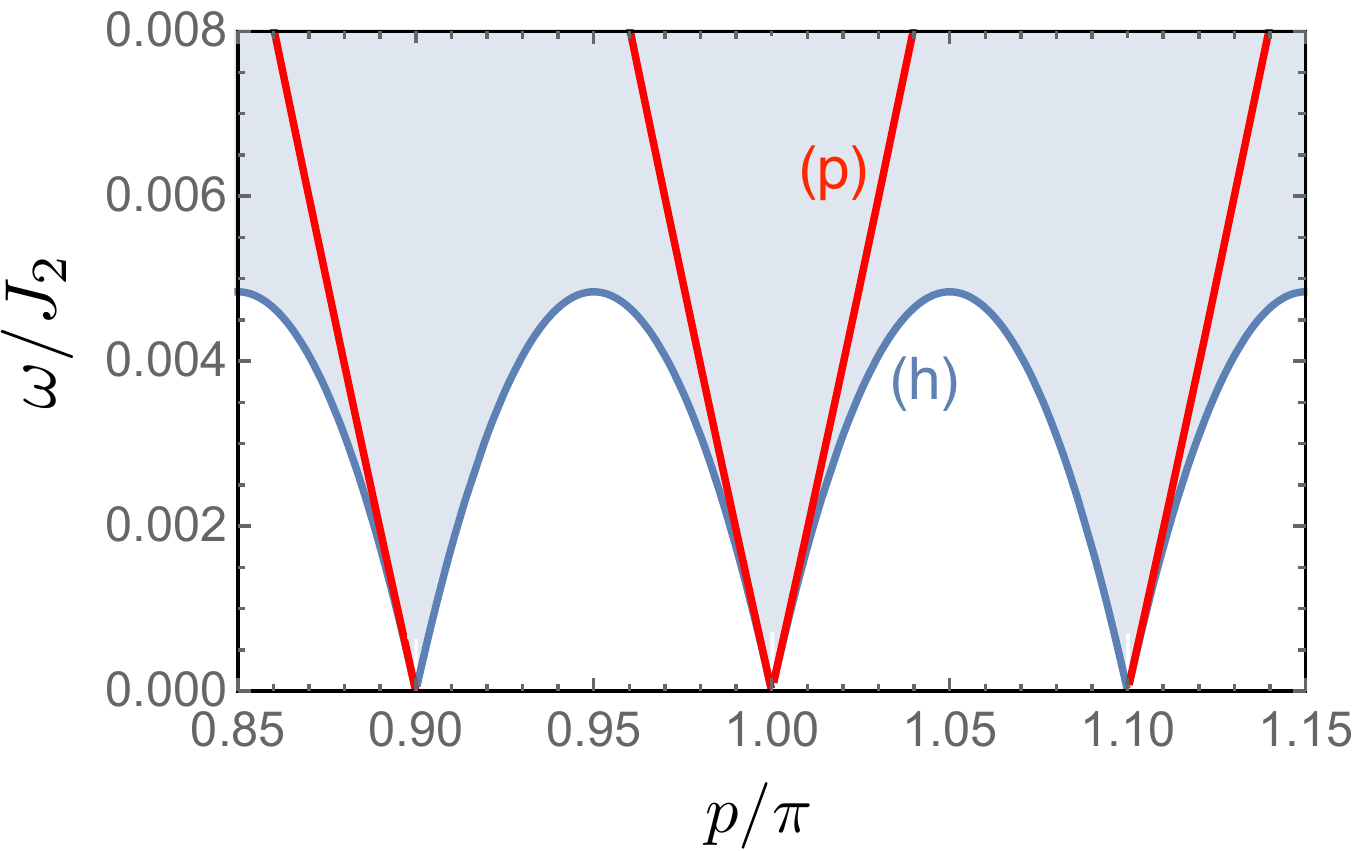}
\end{center}
\caption{Approximate bound-magnon-pair spectrum in the vicinity of $p=\pi$  for $m=0.4$ ($\rho_0=0.05$). The red and blue lines represent the dispersion of   particle-type   and hole-type excitations, respectively. The latter  define the lower threshold of the continuum (shaded area) for excitations with    $\Delta S^z_{\text{tot}}=\pm 2$ (no single magnons). Here we   used $v=\pi \rho_0/M$ for the sound velocity in the limit $\rho_0\ll1$.}
\label{fig:liebmodes}
\end{figure}

The spectrum of the Tonks-Girardeau gas has two types of elementary excitations: a particle-type excitation, which is the 1D analogue  of the Bogoliubov quasiparticle in a superfluid, and a hole-type excitation \cite{LiebLiniger1963,Lieb1963}.  The corresponding dispersion relations are illustrated in Fig. \ref{fig:liebmodes}. For  $|p-\pi| \to 0$, the dispersions of particle-type and hole-type modes become linear, with a slope that defines the sound velocity in the TL model in Eq. (\ref{TLmodel}). In the limit $\rho_0\to0$, the velocity approaches the Fermi velocity $v\to v_F=\pi \rho_0/M$. Moreover,   multiple  ``replicas''  of particle-type  and hole-type  dispersions   are generated by adding particle-hole   excitations with zero energy and momentum $2\pi \rho_0 n$, with $n\in\mathbb Z$.   The gapless excitations at $p-\pi=2\pi  \rho_0 n$ correspond to the series of harmonics in the bosonization formulas for bosons \cite{Haldane1981,Cazalilla2004}. However, the linear-dispersion approximation is valid only at energy scales $\omega\ll v_F \rho_0\sim \rho_0^2/M$. For general $p$, the   dispersion of the hole-type excitation determines the  lower threshold of the continuum for excitations that create  or annihilate  a bound magnon pair. The picture here is qualitatively similar to the spectral functions for the   Lieb-Liniger model \cite{Caux2006,Khodas2007,Imambekov2008}. 

Going beyond the approximations  in Eq. (\ref{Hbcontinuum}), the effective model for bound magnon pairs at low but finite density can be mapped to spinless fermions with a momentum dependent scattering amplitude $V_{b}(p,p',q)$. In fact,  the eigenstates of the hard-core boson Hamiltonian  are in one-to-one correspondence with those of a Fermi system with the same interaction potential \cite{Girardeau1960}. In this case, we expect the nonlinear lower threshold in Fig. \ref{fig:liebmodes} to remain qualitatively valid but deviate from the quadratic momentum dependence  implied by the Galilean invariance of Eq. (\ref{Hbcontinuum}). The gapless points at momenta $|p-\pi|=2\pi \rho_0 n$, $n\in\mathbb Z$,  are  still  fixed by the density of bound magnon pairs. For reference, one can think of the XXZ spin chain for which the relation to the Lieb-Liniger gas of bosons in a scaling limit can be made exact \cite{GaudinBook}.

\begin{figure}[t]
\begin{center}
\includegraphics[width=0.8\columnwidth]{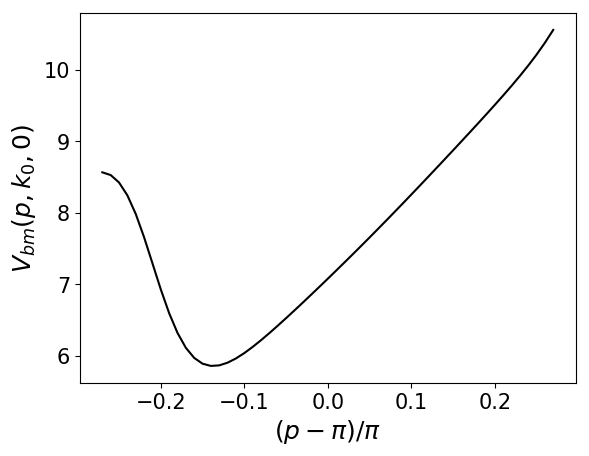}
\end{center}
\caption{Interaction potential $V_{b\text{-}m}(p,k_0,0)$ (with momentum transfer $q = 0$) of a magnon with momentum $k_0$ (corresponding to the minimum of the magnon band) and a bound magnon pair with momentum $p$.}
\label{fig:V}
\end{figure}

Finally, we must consider the interaction between magnons and bound magnon pairs:
\be
H_{b\text{-}m}=\frac{1}{L}\sum_{p,k,q}V_{b\text{-}m}(p,k,q)b^\dagger_{p+q}b^{\phantom\dagger}_pa^{\dagger}_{k-q}a^{\phantom\dagger}_{k}.\label{bmeffectiveinter}
\ee
The relative wave function of magnons and bound magnon pairs also obeys a hard-core constraint in the sense that the single magnon cannot occupy the same position as one of the flipped spins in the bound state, cf. Eq. (\ref{boundstates}). The regular part of the scattering amplitude is given by
  \be
V_{b\text{-}m}(p,k,q)=L\langle b,p+q;k-q| \delta H_{b\text{-}m}(p,k) |b,p;k\rangle,\label{Vbm}
\ee
where $ \delta H_{b\text{-}m}(p,k)=H-E_0-\mc E_b(p)-\varepsilon(k)$. The complete effective Hamiltonian in the low-density limit is 
\be
H_{\text{eff}}=H_b+H_m+H_{b\text{-}m}.
\ee
One immediate  effect of the pair-magnon interaction is to renormalize  the magnon dispersion once there is a finite density of pairs in the ground state.  Within a Hartree-Fock approximation, the renormalized magnon gap is $\tilde \Delta\approx \Delta+\rho_0 V_{b\text{-}m}(\pi,k_0,0)$. For repulsive pair-magnon interactions,  $ V_{b\text{-}m}(\pi,k_0,0)>0$, the effective single magnon dispersion $\tilde \varepsilon(k)$ can remain gapped despite the lowering  of the magnetic field, cf. Eq. (\ref{magnondispersion}). Furthermore,  the interaction of a single magnon with the low-energy  modes of the 1D liquid of bound magnon pairs is important to interpret  the lower threshold of the DSF for excitations with $\Delta S_{\text{tot}}^{z}=\pm1$, as we shall discuss  in  \edit{Sec.} \ref{sec:mim}. For this reason, we have calculated the scattering amplitude in Eq. (\ref{Vbm}). The detailed calculation is presented in   App \ref{app:Vbm}. Figure \ref{fig:V} illustrates the dependence of  the two-particle scattering amplitude on the momentum of the incoming bound magnon pair. We expect this nonuniversal interaction potential to be strongly renormalized in the case of a finite density of bound states. For instance, we find that the result is sensitive to the wave function of the bound states $\Psi(p,r)$ (see \edit{App.}  \ref{app:Vbm}). Nonetheless, the result in Fig. \ref{fig:V} indicates that the magnon-pair interaction is remarkably strong and momentum dependent for the range of parameters in which we are interested. 

\section{Edge singularities\label{sec:mim}}

Spectral functions of critical 1D systems exhibit power-law singularities along special lines in the $(q,\omega)$ plane that determine the thresholds of the support at finite energies. These edge singularities can be described by effective mobile impurity models,  first put forward in Ref. \cite{Pustilnik2006} and reviewed in \cite{Imambekov2012}. The exponents of the edge singularities are nonuniversal, as they depend on phase shifts for the scattering between finite-energy elementary particles and the low-energy modes of the TL liquid. These phase shifts can be either fixed exactly for integrable models \cite{Pereira2008}, or    expressed in terms of phenomenological relations for generic models \cite{Imambekov2009}. 

Here we are interested in the edge singularities of DSFs in the spin-nematic phase. Identifying these singularities will be useful to interpret the numerical  results in \edit{Sec.} \ref{sec:numerics}. Even though the edge exponents  are nonuniversal and depend on phenomenological parameters, we shall take advantage of the low-density limit $\rho_0\ll 1$ to simplify our discussion.

\subsection{Excitations with $\Delta S^z_{\text{tot}}=\pm1$}
\label{sec:edge1}

We start with  the edge singularities that involve single magnon excitations. As discussed in \edit{Sec.} \ref{sec:spectrum}, magnons are gapped particles that interact with the 1D Bose liquid of bound magnon pairs. Let us first discuss the excitations with $\Delta S^z_{\text{tot}}=-1$. The lowest-energy excited states that couple to the ground state via the operator $S_j^-$ contain a magnon with   energy close  to the renormalized gap $\tilde\Delta$. Thus, we can represent the spin-flip operator in the field theory as
\bea
S_{j=x}^-&\sim& \frac1{\sqrt L}\sum_{k}e^{-ikx}a^\dagger_k\nonumber\\
&\approx &e^{-ik_0x}a^\dagger(x),
\eea
where $a^{\dagger}(x)=L^{-1/2}\sum_{|q|\ll k_0}e^{-iqx}a^\dagger_{k_0+q}$ represents the slowly varying components of the magnon with momentum near $k_0$.

In terms of the slowly varying fiels $a(x)$  we can approximate the corresponding term in the effective Hamiltonian as
\be
\label{Heffa}
    H_m \approx \int dx \, a^{\dag}\left( \tilde{\Delta} - \frac{\partial_x^2}{2\tilde{m}} \right)a, 
\ee
where $\tilde{m}$ is the effective mass for magnons with momentum $k\approx k_0$.  The magnon-magnon interaction is dropped because we only consider single-magnon  configurations. The ground state of the effective field theory is a vacuum of magnons, $a(x)\ket{\Psi_0} =0$.

Recall that all the particles are hard-core bosons. To take care of the corresponding scattering phase shift in the effective description, we switch to a ``fermionic'' representation  by attaching Jordan-Wigner strings to the field operators as done in Eq. (\ref{JWb}) for the $b$ fields. Similarly, for the magnons we take 
\bea
a(x)&=&a_F(x)\cos\left[\pi \int_{-\infty}^x dx'\, b^\dagger(x')b(x')\right].
\eea  
The density operators are invariant under this transformation. The effective Hamiltonian in terms of $a_F$ and $b_F$ has the same form as in Eqs. (\ref{Heffb}) and (\ref{Heffa}). However, the hard-core constraint is now automatically  satisfied and the $s$-wave scattering amplitudes have no effect on the   $a_F$ and $b_F$ particles. 

To obtain the  mobile impurity model, we project the effective Hamiltonian onto a magnon sub-band with  cutoff   scale $\Lambda\ll Mv_F^2$ and bosonize the low-energy modes. The uniform part of the density of bound magnon pairs becomes
\be
b^\dagger(x)b(x)\sim \rho_0+\frac{1}{\sqrt \pi}\partial_x\phi(x).
\ee
The spin-lowering operator is then represented by \be
S_{j=x}^-\sim e^{-ik_0x} a^\dagger_F(x)\cos\left[\pi \rho_0 x+\sqrt{\pi }\phi(x) \right].
\ee
The single magnon plays the role of the mobile impurity in the model   $  H_{\text{imp}}=\int dx\, \mc H_{\text{imp}}(x)$ with Hamiltonian density 
 \bea
 \label{Himp}
\mc H_{\text{imp}}&=&  \frac{vK}{2}(\partial_x\theta)^2+\frac{v}{2K}(\partial_x\phi)^2+a_F^\dagger\left(\tilde{\Delta}-\frac{\partial_x^2}{2\tilde{m} }\right) a^{\phantom\dagger}_F\nonumber\\
&&+\frac{ v}{\sqrt\pi}\left( \gamma_1 \partial_x\phi + \gamma_2 \partial_x\theta \right)a^{\dagger}_F a^{\phantom\dagger}_F,
\eea
where    $\gamma_{1,2}$  are phenomenological coupling constants. They descend from the interaction $V_{b\text{-}m}$ and represent the  coupling between  the magnon with momentum $k_0$ and   the local density or current of bound magnon pairs. For now, let us consider the asymptotic low-density limit,  which corresponds to putting $K=1$; let us also set $\gamma_{1,2} =0$. In this limit we can calculate the time-dependent correlation
\bea
 \langle S^+_{x}(t)S^-_{0}(0)\rangle &\sim &e^{i(k_0\pm \pi \rho_0)x}\langle a^{\phantom\dagger}_F(x,t)a_F^\dagger(0,0)\rangle \nonumber\\
&&\times \langle e^{\pm i\sqrt \pi \phi(x,t)}e^{\mp i\sqrt \pi \phi(0,0)} \rangle\nonumber\\
&\sim&\frac{e^{i(k_0\pm \pi \rho_0)x -i\tilde \Delta t}\,G(x,t)}{|x^2-v^2t^2|^{1/4}},\label{twomomenta}
\eea
where \be
G(x,t)=\sqrt{\frac{\tilde{m} }{2\pi i t}}e^{i\tilde{m}  x^2/(2t)}\label{propagator}
\ee
is the   propagator for the free particle with mass $\tilde{m}$. Note the momentum shift $\pm \pi \rho_0$ in the spatial oscillation of the  correlation in Eq. (\ref{twomomenta}). This means that the minimum energy in the corresponding DSF, $S^{+-}(q,\omega)$, does not occur at the momentum $k_0=\arccos(|J_1|/4J_2)$. Instead, the edge singularity is split into two. Taking the Fourier transform with either momentum yields the threshold behavior:\bea
&&S^{+-}(q=k_0\pm\pi \rho_0,\omega)\nonumber\\
&\sim& \int  dx dt\, e^{i\omega t}  e^{-i(k_0\pm \pi \rho_0)x}\langle S^+_{x}(t)S^-_{0}(0)\rangle\nonumber\\
&\sim&\Theta(\omega-\tilde\Delta)(\omega -\tilde\Delta)^{-1/2}, \label{mu+-ld}
\eea
where $\Theta(x)$ denotes the Heaviside step function. Allowing for $K\neq 1$ or $\gamma_{1,2} \neq 0$  leads to a similar threshold behavior but the exponent  of the power law changes as an effect of interactions. The full  effective field theory result reads
$S^{+-}(q=k_0\pm\pi \rho_0,\omega) \sim \Theta(\omega-\tilde\Delta)(\omega -\tilde\Delta)^{\mu^{+-}_{\pm}}$
with the exponent
\be
\label{mu+-}
\mu^{+-}_{\pm} =  \frac{1}{2} \left(\frac{\gamma_1 \sqrt{K}}{\pi}\right)^2 + \frac{1}{2}\left(\sqrt{K}\mp \frac{\gamma_2}{\pi \sqrt{K}} \right)^2 - 1,
\ee
(see \edit{App.} \ref{app:exponents}).
The important feature to note is the asymmetry between $\mu^{+-}_{\pm}$ (where the lower index corresponds to the threshold at $q = k_0 \pm \pi \rho_0$, respectively) when the coupling constant $\gamma_2$ is nonzero.  
 
Now let us discuss the excitations with $\Delta S^{z}_{\text{tot}}=+1$. Since we cannot annihilate a single magnon in the ground state, the simplest possible excitation with  the proper quantum number must involve the creation of a magnon and annihilation of a bound magnon pair. Thus, we represent the spin-raising operator by
\bea
\label{eq:projSp}
S^+_{j=x}&\sim& e^{-ik_0x}(-1)^xa^\dagger(x)b(x),
\eea
where the factor of $(-1)^x$ reflects the momentum $\pi$ carried by the low-energy bound magnon pair, cf. Eq. (\ref{bcontinuumpi}).  We bosonize the pair operator as \cite{Haldane1981,Cazalilla2011}
\be
    b(x)\sim e^{-i \sqrt{\pi} \theta(x)}.
\ee
Together with the Jordan-Wigner string for the magnon operator, this leads to the representation for the time-dependent correlation
\begin{align}
  &\langle  S^{-}_{x}(t)S^+_0(0) \rangle \sim  e^{i(\pi + k_0 \pm \pi \rho_0)x}\\
& \times \langle e^{i\sqrt{\pi}[\theta(x,t) \pm \phi(x,t)]} a_F(x,t)a^{\dag}_F(0,0)e^{-i\sqrt{\pi}[\theta(0,0) \pm \phi(0,0)]}\rangle\nonumber.
\end{align}
The computation of the threshold behavior of the structure factor follows the same lines as before and leads to the result $S^{-+}(q = \pi + k_0 \pm \pi \rho_0,\omega) =   \Theta(\omega-\tilde\Delta)(\omega -\tilde\Delta)^{\mu^{-+}_{\pm}}$ with exponent
\be
\label{mu-+}
\mu^{-+}_{\pm} =   \frac{1}{2} \left(\frac{1}{\sqrt{K}}+
\frac{\gamma_1 \sqrt{K}}{\pi}\right)^2 +\frac{1}{2}\left(\sqrt{K}\mp \frac{\gamma_2}{\pi \sqrt{K}} \right)^2 - 1.
\ee
This exponent vanishes in the low-density limit: $\mu^{-+}_{\pm} = 0$ when $K=1$ and $\gamma_{1,2} = 0$. The threshold behavior in this case is thus strongly dependent on interaction effects. In particular, note that the $\gamma_1$,  which encodes density-density interactions between the magnon and the bound state, is expected to be positive for repulsive interactions. As such, it cannot render the singular behavior divergent. The $\gamma_2$ interaction, which encodes the asymmetry in the coupling of the magnon to the right and left moving modes of the Luttinger liquid, can lead to divergent behavior in one of the thresholds at, say, $q = \pi + k_0 +\pi \rho_0$ but that would imply that the structure factor is convergent at the other threshold at $q = \pi + k_0 -\pi \rho_0$.

\subsection{Excitations with $\Delta S^z _{\text{tot}} = \pm 2$}
 \label{sec:edge2}
We continue    by considering the two-particle structure factors   $S^{\pm\pm \mp\mp}(q,\omega)$. These are associated with operators flipping two spins on neighboring sites for which the simplest low-energy excitation in the effective field theory is a single bound state created on top of, or annihilated from, the condensate of bound magnon pairs.
 We represent the two-spin-flip operators \edit{therefore}  as
\bea
 S^{-}_{x}S^{-}_{x+1} &\sim& (-1)^xb^{\dag}(x),\label{eq:b-particle}\\
 S^{+}_{x}S^{+}_{x+1} &\sim& (-1)^xb(x).\label{eq:b-hole}
\eea
The representation for the structure factors becomes

\be
  S^{++--}(q,\omega)  \sim  \int dx dt e^{i(\omega t-qx)} (-1)^x\langle b(x,t)b^{\dag}(0,0)\rangle,
\ee  
\be
   S^{--++}(q,\omega)  \sim \int dx dt e^{i(\omega t-qx)} (-1)^x\langle b^\dag(x,t)b(0,0)\rangle.
\ee

In this language, these two-spin DSFs are equivalent to the particle and hole spectral function of hard-core bosons. 
The analogues of these functions in the integrable XXZ model would be the transverse structure factors $S^{\pm \mp}(q,\omega)$. Even in the simplest case where the mapping to free fermions is exact, such as \edit{for} the XX chain, these are not easy to calculate because of the non-trivial appearance of the Jordan-Wigner string. The function $S^{-+}(q,\omega)$ has been computed numerically for XXZ by Bethe ansatz based techniques \citep{Caux2005}. \edit{Field theory methods similar to those in the previous subsection can give considerable insight into these functions as demonstrated in Refs. ~\cite{Karimi2011,Imambekov2012}.} 

In contrast with the DSFs for $\Delta S^z=\pm1$, the lower threshold of the support now extends down to zero frequency at $q=\pi +2\pi \rho_0 n$, $n\in\mathbb Z$. At these values of $q$  and for $\omega\to0$, bosonization predicts  $S^{\pm\pm \mp\mp}(\pi +2\pi \rho_0 n,\omega)\sim \omega^{2(n^2K-1)+1/(2K)}$. Thus,   divergent behavior is  expected  at  $q=\pi$. Away from the gapless points, the nonlinear threshold is determined by hole-type excitations as discussed in Sec. \ref{subsec:beyond}. The phenomenological parameters of the effective mobile impurity model can vary continuously along this threshold and depend on the interaction between the hole-type excitation and the low-energy modes of the TL liquid.

\section{Numerical results\label{sec:numerics}}
In this section, we show the numerical results of DSFs for the frustrated ferromagnetic spin chain described by the Hamiltonian in Eq.~(\ref{model}). In order to obtain the time-dependent correlations, we have used the adaptive tDMRG method developed by Feiguin and White \citep{FeiguintDMRG}. This method is most efficient to investigate chains with nearest neighbor interactions. For systems with short-range interactions, such as narrow ladders and the $J_1 \text{-} J_2$ Heisenberg chain, it is convenient to use the supersite version of the adaptive tDMRG. The central idea of this version is to combine single sites into a supersite such that the Suzuki-Trotter decomposition of the time evolution operator can be applied exactly in the non-renormalized DMRG sites.

To investigate the DSFs, we have considered open chains with system size $L=400$. All the numerical results were obtained by setting $J_1=-1$ and $J_2=1$. We take magnetizations $m=0.2$ and $m=0.4$ to represent  the SDW and and quadrupolar nematic regimes, respectively. Equivalently, these magnetizations correspond to magnetic fields $h=0.95$ ($m=0.2$) and  $h=1.238$ ($m=0.4$).  We  determine the single-magnon gap using DMRG by computing the energy differences\be
\Delta E_\pm(m,L) =E_{gs}(M=mL\pm 1)-E_{gs}(M=mL),
\ee
where $E_{gs}(M)$ is the lowest energy in the sector with $S^z_{\text{tot}}=M$ of the chain with length $L$. The magnon gap is given by \be
\tilde \Delta(m)=\lim_{L\to\infty} \Delta E_+(m,L)=\lim_{L\to\infty} \Delta E_-(m,L).
\ee
We obtain the values $\tilde\Delta\approx 0.13$ for $m=0.4$ and $\tilde\Delta\approx 0.14$ for $m=0.2$.

In order to compute the two- and four-point time-dependent correlations, we kept up to $400$ states to represent the restricted Hilbert space of each DMRG block. Typically, the error associated with the truncation procedure is smaller than $10^{-7}$. The time evolution was carried out with second order Suzuki-Trotter decomposition of the time evolution operator with time step $\delta t=0.05$. As is well known, the Trotter error  
depends on the order $n$  of the Suzuki-Trotter decomposition. For the order $n$, the Trotter error is of order $\left(\delta t\right)^{n+1}$. It is worth to mention that we have checked our code by reproducing results for  the integrable XXZ chain \citep{Caux2006,Klauser2011} and for the $J_1\text{-}J_2$ Heisenberg chain at zero magnetization \citep{Ren2012}. 

As shown in Eqs.~(\ref{onespin}) and (\ref{twospin}), the DSFs can be acquired by performing the Fourier transform  of the time-dependent correlations computed  in the space and time domains.  
Since we only have numerical results for finite time, the temporal Fourier transforms were performed in the time interval $-t_{\text{max}}<t<t_{\text{max}}$, where $t_{\text{max}}$ is the maximum time obtained by  tDMRG. In our computations we have considered   $t_{\text{max}}$ in the interval $t_{\text{max}}\in[35,60]$. In the following, we present  our numerical results and interpret them in terms of the effective theory of gapped magnons and gapless bound magnon pairs.

\subsection{One-spin structure factors}

\begin{figure}[t]
\begin{center}
\includegraphics[width=0.95\columnwidth]{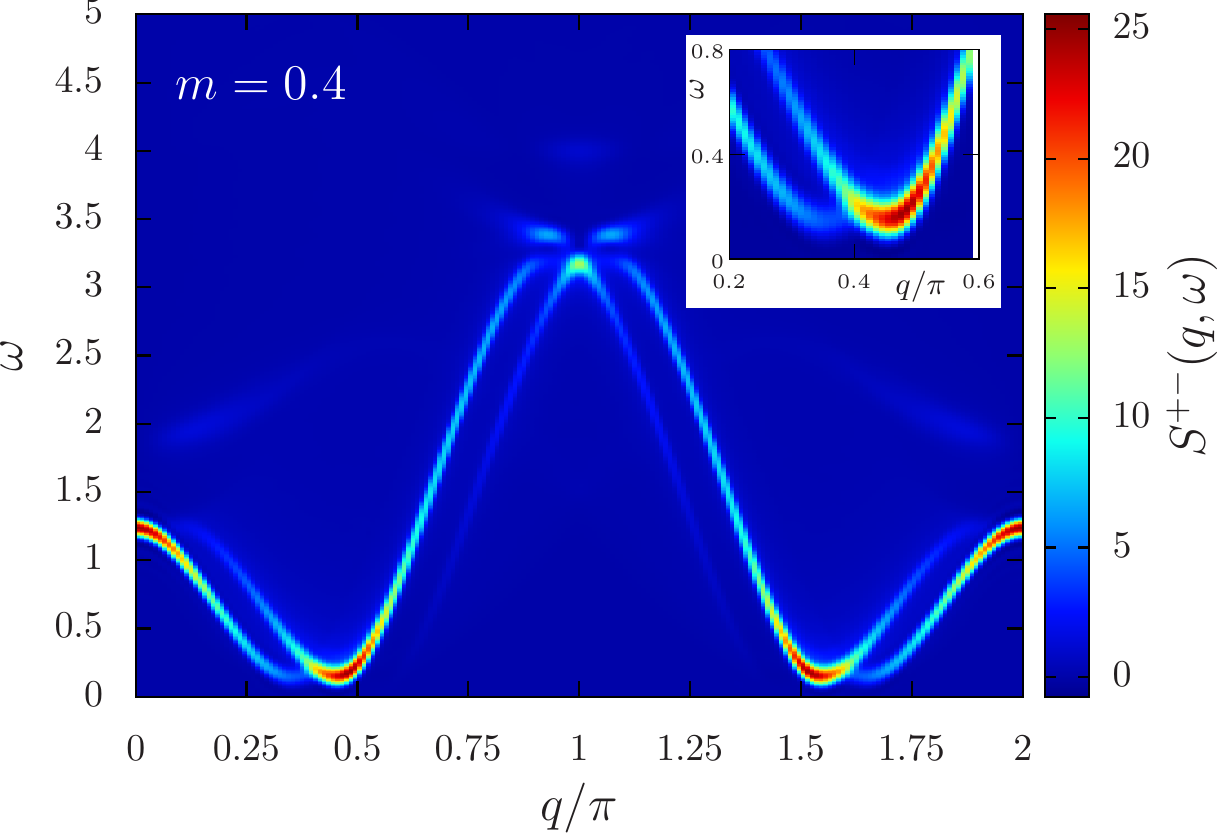}
\end{center}
\caption{The dynamical structure factor $S^{+-}(q,\omega)$  in the nematic phase with $J_2=-J_1=1$ and $m=0.4$ computed by tDMRG. The lower threshold is associated with the creation of a single magnon in the effective theory. 
The two copies  at low energies are associated with momentum shifts $\pm \pi\rho_0$  due to the hard-core repulsion between the magnon and the pairs in the condensate. The inset shows a zoom-in on the minima located in the interval $0.2\pi\le q\le 0.6\pi$. }
\label{fig:Spm}
\end{figure}

\begin{figure}[t]
\begin{center}
\includegraphics[width=0.95\columnwidth]{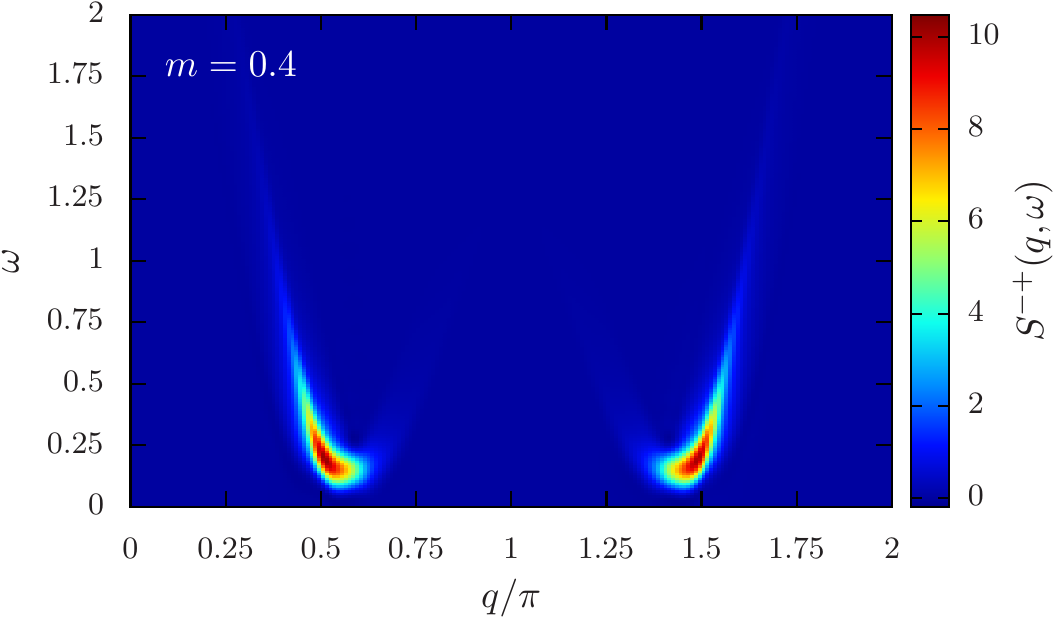}
\end{center}
\caption{The dynamical structure factor $S^{-+}(q,\omega)$.    The lower threshold is associated with the creation of a magnon and the annihilation of a bound magnon pair. In this case only one threshold  is visible. 
}
\label{fig:Smp}
\end{figure}

The tDMRG result for the structure factor $S^{+-}(q,\omega)$ in the nematic phase is presented in Fig.~\ref{fig:Spm}. Note that the intensity is concentrated on curves that (at least qualitatively) follow the shape of the magnon dispersion. The lower edge of the support are the thresholds for which the effective theory in terms of a single gapped magnon interacting with the bound-state condensate was formulated. The magnetization $m=0.4$ corresponds to a bound-state density of $\rho_0 = 0.05$ according to Eq.~(\ref{rho0}).

Above the saturation field, the magnon dispersion has a minimum at $k_0=\cos^{-1}(1/4)\approx 0.42\pi$. According to the discussion in Sec.~\ref{sec:edge1}, we expect to find minima 
of the thresholds at $q = k_0 \pm \pi \rho_0 \approx ( 0.42 \pm 0.05)\pi$ \edit{ for $0 < q <\pi$ (the domain $\pi < q <2 \pi$  can be obtained by reflection over $q = \pi$).} In Fig~(\ref{fig:Spm}), we see that this is indeed close to where the threshold minima are found. \edit{We  note that there is a clear asymmetry between the two minima, which we interpret as due to the momentum dependence of the magnon-pair interaction which accounts for a nonzero $\gamma_2$ coupling in the effective model of Eq. (\ref{Himp}).} Strictly speaking, the threshold exponents  only make sense in the thermodynamic limit and with infinite energy resolution. However, we expect that the thresholds characterized by more negative exponents and thus stronger divergences appear with greater intensity in the tDMRG results, as indeed verified in simpler models \cite{Pereira2008}.  In fact, for $\gamma_2 >0$, the result in Eq. (\ref{mu+-}) predicts $\mu^{+-}_{+}<\mu^{+-}_{-}$, thus  a brighter  threshold at $q = k_0 + \pi \rho_0$ than at  $q = k_0 - \pi \rho_0$.   

\begin{figure}[t]
\begin{center}
\includegraphics[width=0.95\linewidth]{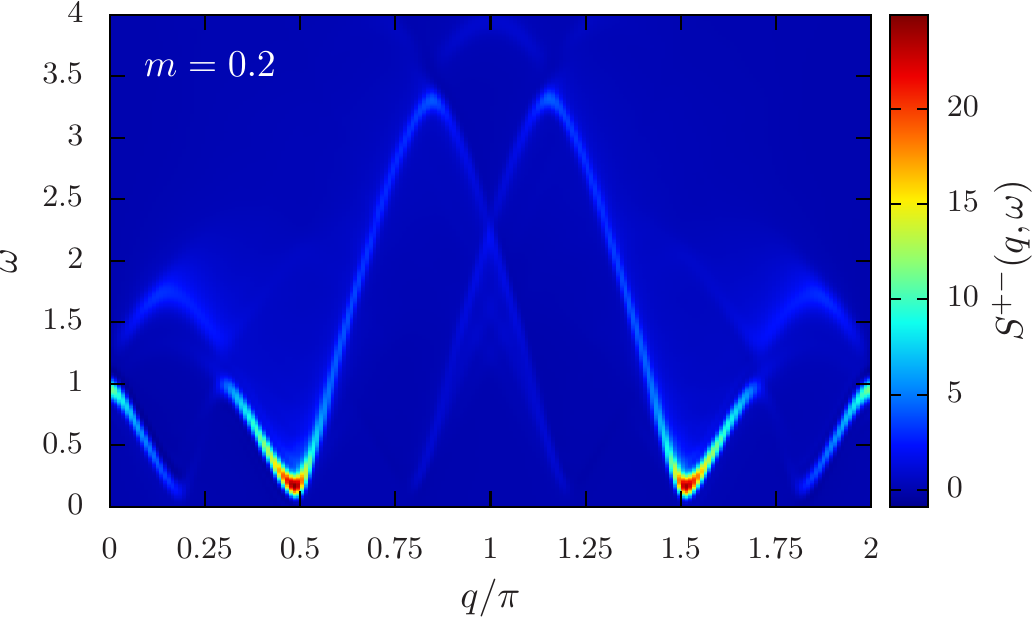}
\includegraphics[width=0.95\linewidth]{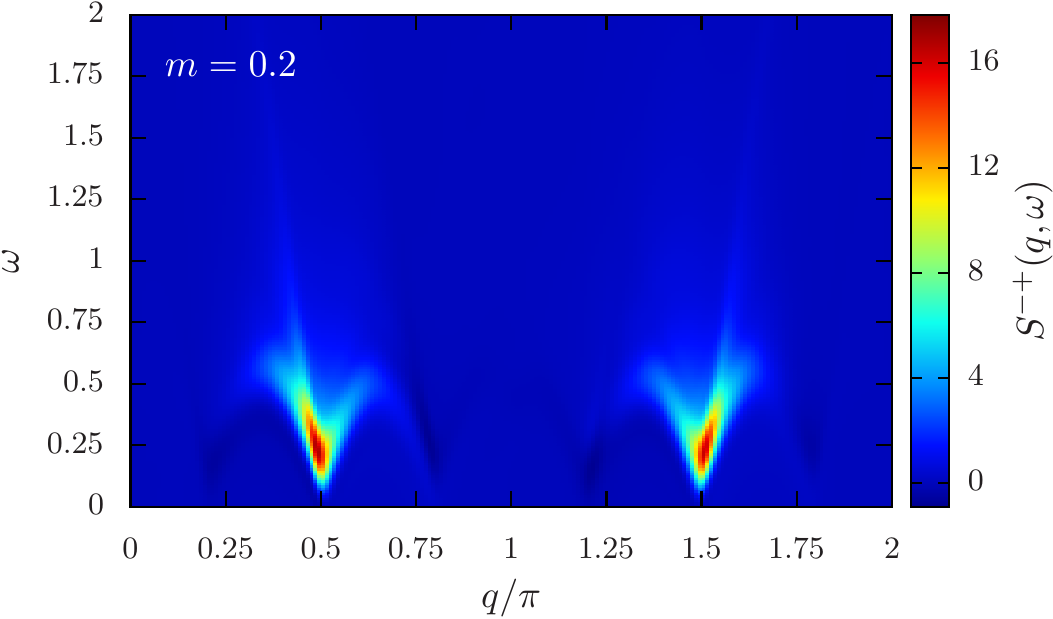}
\end{center}
\caption{One-spin-flip dynamical  structure factors in the SDW phase with $m=0.2$. Compared to the spin nematic phase, we see a larger splitting  $\pm  \pi \rho_0$ in $S^{+-}(q,\omega)$ (top). In $S^{-+}(q,\omega)$ (bottom),  the lower threshold  associated with    hole-type excitations of  the bound-state condensate is now visible.}
\label{fig:S1-sdw}
\end{figure}

Let us turn attention to the DSF $S^{-+}(q,\omega)$, associated with the creation of a magnon in conjunction with the annihilation of a bound magnon pair. The tDMRG result is shown in Fig.~\ref{fig:Smp}. In contrast with  $S^{+-}(q,\omega)$, one immediate observation is that there is no clear sign of the two replicas of the threshold. This is exactly what should be expected in light of the expression for the exponent $\mu^{-+}_{\pm}$ in Eq.~(\ref{mu-+}): At one threshold we find divergent behavior visible as a high intensity peak in   the tDMRG data, while the other threshold has a positive  exponent   corresponding to a  vanishing   intensity at the threshold, \edit{which  means the threshold becomes invisible in the tDMRG data.} To see whether we indeed find consistent results with the foregoing discussion, let us   check the expected position and qualitative behavior of the threshold. First of all, note that the momenta of the minima are now at $\pi + k_0 \pm \pi \rho_0$, in agreement  with the effective theory. Let us discuss the minima in the domain  $\pi < q < 2\pi$ (the domain $0<q<\pi$ can be obtained by reflection in $q=\pi$). A comparison of the exponents $\mu^{-+}_{\pm}$ and $\mu^{+-}_{\pm}$ shows that  the parameter $\gamma_2$ gives a stronger divergence at the  momentum associated with the same shift for  $S^{-+}(q,\omega)$ as for $S^{+-}(q,\omega)$. This means that, for $\gamma_2>0$, the visible threshold in the tDMRG data is expected at $q = \pi+k_0+\pi \rho_0 \approx  1.47\pi$. Inspection of Fig.~\ref{fig:Smp} shows that this is indeed the case (the invisible threshold would be around $q \approx 1.37\pi$). Furthermore, one can inspect what happens with the intensity at the threshold away from the minimum. The intensity shows a stronger increase as one goes to $q \gtrsim \pi + k_0 + \pi \rho_0$ in Fig.~\ref{fig:Smp}. In Fig.~\ref{fig:Spm}, we see that this also happens for $q \gtrsim  k_0 + \pi \rho_0$ in the case of $S^{+-}(q,\omega)$. This asymmetry can be attributed  to the dependence of the effective pair-magnon interaction on the magnon momentum $k\neq k_0$.

\begin{figure}[t]
\begin{center}
\includegraphics[width=0.75\linewidth]{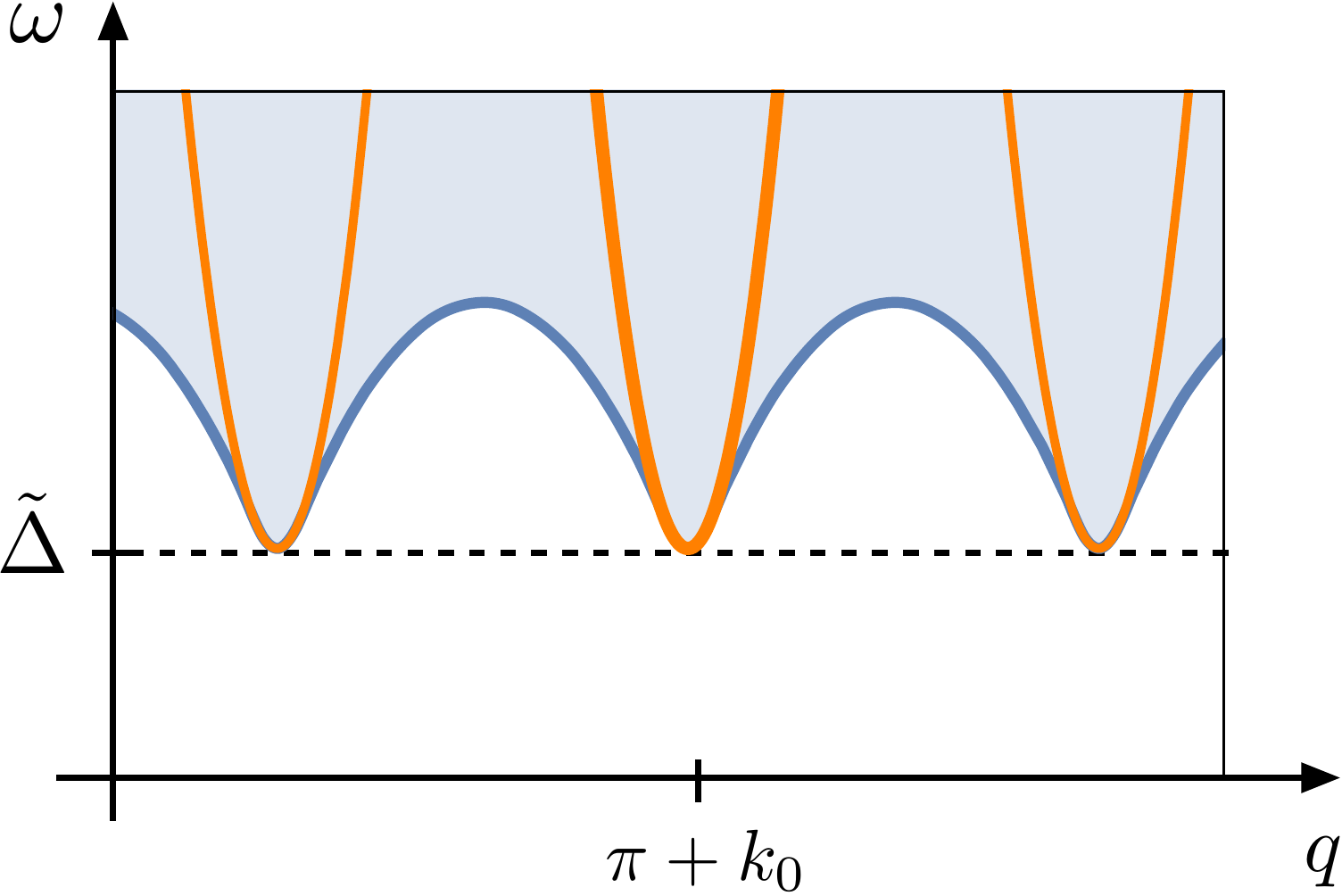}
\end{center}
\caption{Schematic representation  of the two-particle continuum $E(k,q)=\varepsilon(k)+\mc E_b(q-k)$. Near $q=\pi+k_0+2\pi\rho_0n$, with $n\in\mathbb Z$, the lower threshold  coincides with the renormalized magnon dispersion. As we deviate from these points and the magnon velocity increases beyond the sound velocity $v$ of the condensate, the lower threshold becomes defined by a magnon and a hole-type excitation with the same velocity, such the energy of the two-body state is minimized. }
\label{fig:loweredge1magnon}
\end{figure}

For comparison, we have also computed the one-spin structure factors in the SDW phase with magnetization $m=0.2$ (Fig.~\ref{fig:S1-sdw}). In this case, the density of bound states is  $\rho_0 = 0.15$, leading to an appreciably larger split  $2\pi \rho_0$ of the band minima observed in the DSFs. This higher bound-state density in the SDW is also expected to lead to much stronger deviations from the low-density limit. Nevertheless,   many features of the DSFs   remain qualitatively the same. We recall that there is no true phase transition between the spin nematic and the SDW ``phases'', which are different regimes of the same TL phase. The biggest difference from the nematic phase is visible in the $S^{-+}(q,\omega)$ structure factor. In this case one still sees the minimum at $q = \pi + k_0 + \pi \rho_0$ associated with a gapped magnon  and  a zero-energy bound state. But emanating from this point, one sees arcs defining the lower threshold of the magnon-pair continuum. This lower threshold is distinguished from the magnon dispersion in the momentum range where the magnon velocity [defined from the renormalized dispersion as $\partial_k\tilde{\varepsilon}(k)$] becomes larger than the sound velocity $v$. Figure \ref{fig:loweredge1magnon} illustrates the two-particle continuum constructed from the dispersion relations $\varepsilon(k)$ and $\mc E_b(p)$. If we include the momentum shift \edit{$\pm\pi\rho_0$}, the edge   of this continuum  can be identified with the lower threshold of  $S^{-+}(q,\omega)$ seen in Fig. \ref{fig:S1-sdw}.

\subsection{Two-spin structure factors}

 \begin{figure}[t]
\begin{center}
\includegraphics[width=0.95\columnwidth]{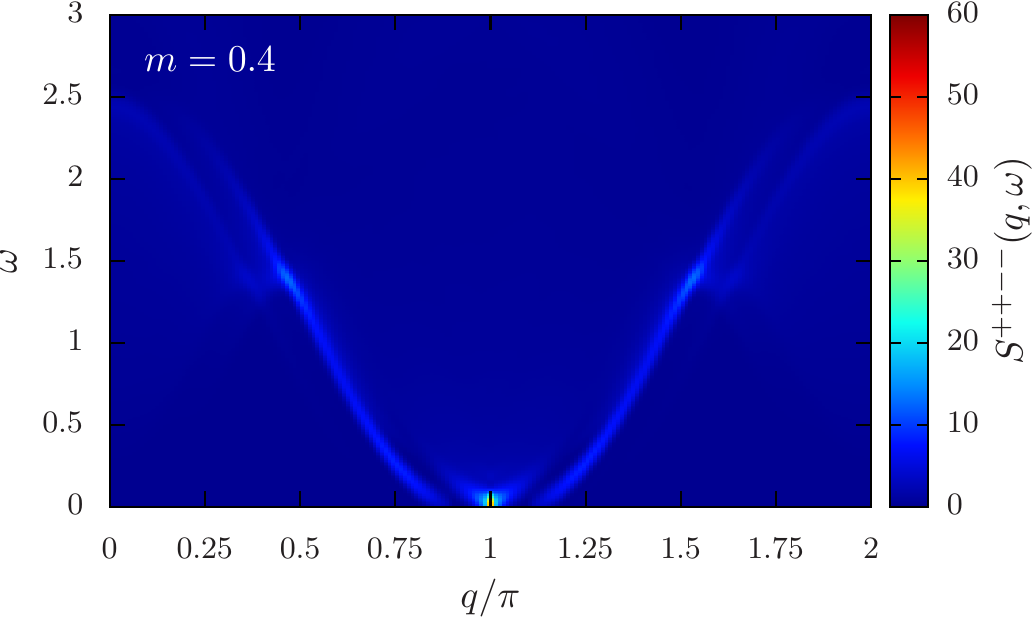}
\includegraphics[width=0.95\columnwidth]{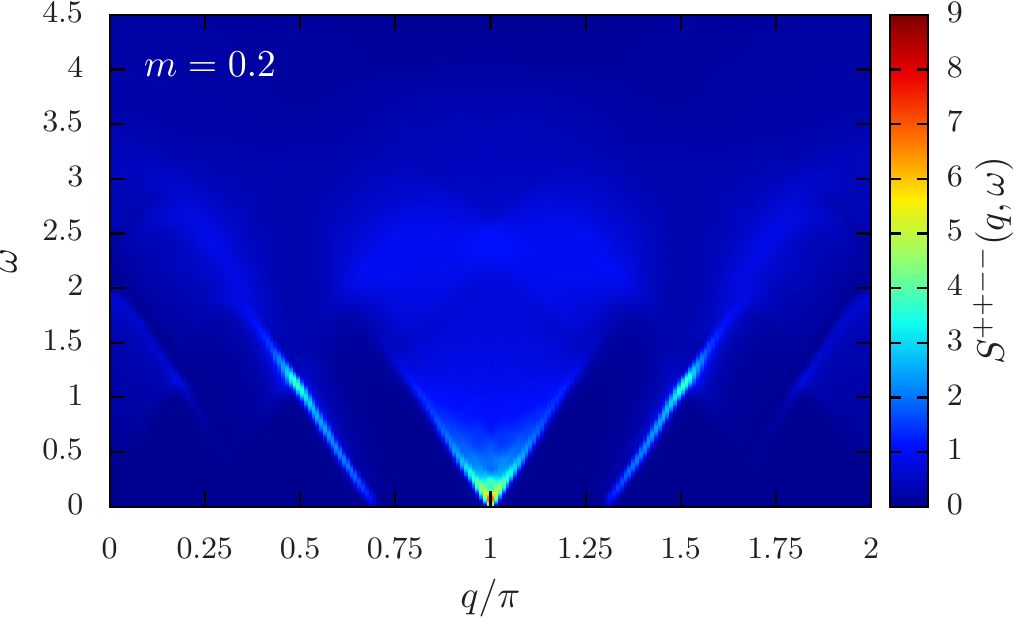}
\end{center}
\caption{Two-spin dynamical structure factors $S^{++--}(q,\omega)$ in the spin nematic phase (top) and SDW phase (bottom). The low-energy spectrum  is associated with   a particle-type excitation  of the condensate.  }
\label{fig:S2}
\end{figure}

\begin{figure}[t]
\begin{center}
\includegraphics[width=0.95\columnwidth]{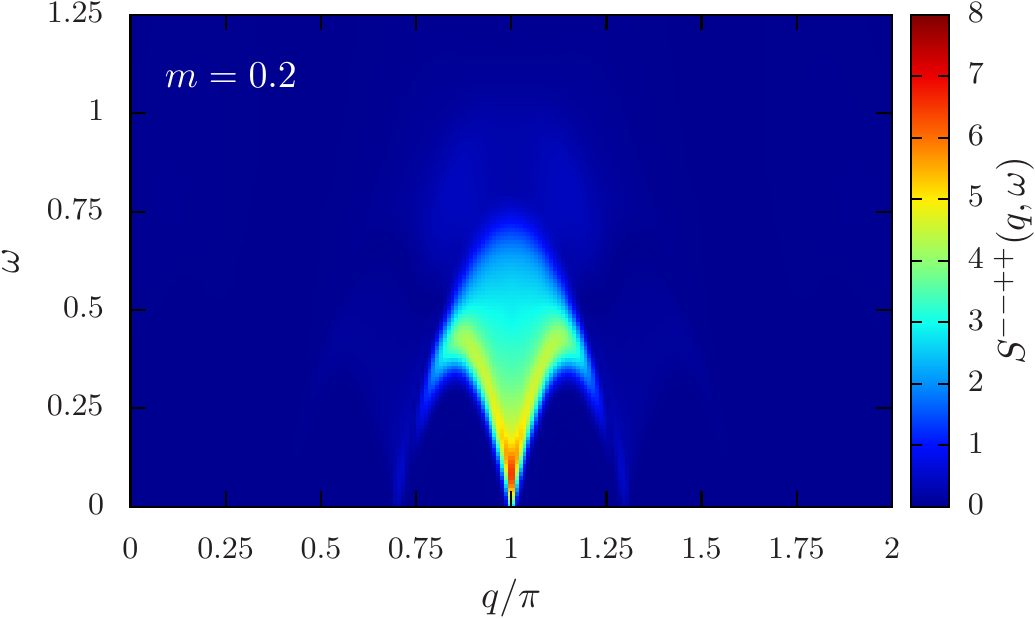}
\end{center}
\caption{Two-spin dynamical structure factor  $S^{--++}(q,\omega)$ in the  SDW phase. The low-energy spectrum is associated with the hole-like excitation of the condensate. }
\label{fig:S2mp}
\end{figure}

The two-spin DSFs $S^{\pm\pm\mp\mp}(q,\omega)$  computed by tDMRG are shown in Figs.~\ref{fig:S2} and \ref{fig:S2mp}. 
\edit{In terms of the effective field theory,} we can view  $S^{++--}(q,\omega)$ as the particle spectral function of the $b$ particles, while $S^{--++}(q,\omega)$ is analogous to   the hole spectral function [cf. Eqs (\ref{eq:b-particle}) and (\ref{eq:b-hole})]. 

The spectral function for   hard-core bosons  is more complicated than for dual fermions due to the Jordan-Wigner string.  Yet, focusing on the function $S^{++--}(q,\omega)$ in Fig. \ref{fig:S2}, we see lines of intensity starting from momentum $q=\pi$ and energy $\omega =0$ as a clear signal of the gapless  dispersion of the particle-type excitation of the condensate, cf. Fig. \ref{fig:liebmodes}. Replicas are found at $q = \pi \pm 2\pi \rho_0$, associated with a particle-type excitation \edit{of momentum $\pi$ dressed by an additional Umklapp-like particle-hole excitation with momentum $\pm 2\pi \rho_0$.} The highest intensity is observed at $q=\pi$ and $\omega\to 0$, where the TL theory predicts a divergent power-law singularity.

Turning to $S^{--++}(q,\omega)$, we have observed that in the nematic phase the spectral weight of this DSF is rather small and highly concentrated near $q=\pi$ and $\omega=0$. This is somewhat expected because $S^{--++}(q,\omega)$ is dominated by the creation of a  hole in the condensate of bound magnon pairs, which can be pictured as a shallow Fermi sea with  a small energy scale ($\sim \rho_0^2/M$) at the low density $\rho_0=0.05$. The concentration at momentum close to $q=\pi$ and low frequencies  implies that our numerical tDMRG result in this case suffers from strong finite-size and finite-time effects. On the other hand, in the SDW phase with $m=0.2$ we observe a clear continuum in the $(q,\omega)$ plane, as shown in Fig. \ref{fig:S2mp}. This result is reminiscent of transverse structure factor  $S^{-+}(q,\omega)$ for the XXZ spin chain computed in Ref.~\onlinecite{Caux2005}.  In contrast with the XXZ chain, where the elementary excitations are single particles and holes in the ground state configuration \edit{of the Bethe ansatz solution}, here the gapless excitations that define the low-energy continuum in $S^{--++}(q,\omega)$ are bound magnon pairs. Despite the different $\Delta S^z$ quantum number of the ``elementary'' excitations,   these two DSFs  are qualitatively similar as both of them  can be interpreted as   hole spectral functions of hard-core bosons.

\section{Conclusion\label{sec:conclusion}}

We have investigated dynamical structure factors of the one-dimensional spin nematic phase using the adaptive time-dependent density matrix renormalization group.  The main features of the excitation spectrum can be understood in terms of gapped single-magnon excitations and a gapless quasi-condensate   of two-magnon bound states, in which elementary particle-type and hole-type excitations carry spin quantum numbers $\Delta S^z=\pm2$ and have nonlinear dispersion.  

The nonzero magnon gap can be discerned in both dynamical structure factors that involve a single spin flip, namely $S^{+-}(q,\omega)$ and $S^{-+}(q,\omega)$. However, the simplest excitations (in the sense of minimum number of elementary particles) are different for these two structure factors. While $S^{+-}(q,\omega)$ is dominated by single-magnon excitations, $S^{-+}(q,\omega)$ involves  the creation of  a magnon and a hole-type excitation.  Remarkably, $S^{+-}(q,\omega)$ shows a doubling of the low-energy threshold due to   momentum shifts $\pm \pi \rho_0$, where $\rho_0$ is the density of bound magnon pairs in the condensate.  This effect is captured by an effective field theory that goes beyond the TL liquid theory by describing the interaction between the single magnon and  the low-energy modes of the condensate, taking into account the hard-core constraints. The asymmetry in the intensity of the doubled threshold in $S^{+-}(q,\omega)$ can be attributed to the momentum dependence of the  effective pair-magnon interaction. This shows that a simple picture of hard-core bosons that neglects the finite-range part of the interaction potentials   is not sufficient to describe the dynamical properties of the spin nematic phase even at rather low values of  $\rho_0$. We have also studied the same structure factors in the SDW regime (at lower magnetization) and found qualitatively similar behavior in the finite-energy spectrum.  One notable feature is that the lower-threshold of the multiparticle continuum, which depends on the dispersion of hole-type excitations of the condensate, becomes clearly visible in $S^{-+}(q,\omega)$  in the SDW regime.  

The dynamical structure factors that involve two spin flips, $S^{++--}(q,\omega)$ and $S^{--++}(q,\omega)$, can be interpreted as spectral functions of bound magnon pairs. In these cases, the lower threshold of the multiparticle continuum extends to zero energy, as the excitations do not necessarily involved gapped magnons. Besides momentum $q=\pi$, gapless points can be seen at $q=\pi\pm 2\pi \rho_0$,  showing a dependence on the density of bound magnon pairs in the condensate. 

\edit{
Several possible directions for future research present themselves. Sticking to the one-dimensional case, an obvious next step would be to study the structure factors at finite temperatures. In principle this is possible by a finite temperature generalization of the tDMRG method \cite{finite-TDMRG-Verstraete} and of the impurity model \cite{2015_Karrasch_NJP_17}. The particular interest would be to extract the NMR spin-lattice relaxation rate connected to the experiments \cite{Orlova2017} on LiCuVO$_4$.  Finally, we envision using the one-dimensional results as a basis for studying two- or three-dimensional structures composed of coupled chains or ladders where the inter-chain coupling is treated using chain mean-field theory. For instance, one can extend on the ideas in Ref. \cite{Starykh2014} in studying the low-energy excitations. Furthermore, the tDMRG data can serve as input for computation of the structure factors in higher dimensions in a random phase approximation similar to Refs. \cite{2011_Bouillot_PRB_83,2017_Blosser_PRB_96}. It  would   be interesting to compare this approach with the results of other approximations for the dynamics of the spin nematic state on the square lattice \cite{2013_Shindou_PRB_87,2015_Smerald_PRB_91,2016_Smerald_PRB_93}.
}

\begin{acknowledgements}
We thank J. C. Xavier for helpful discussions \edit{ and the High Performance Computing Center (NPAD) at UFRN for providing computational resources}. We acknowledge financial support from the Brazilian ministries MEC and MCTIC.
\end{acknowledgements}

%

%


\appendix

\section{Interaction matrix elements\label{app:Vbm}}

Here we show the computation for the pair-magnon interaction $V_{b\text{-}m}(k,p,q)$ with $q=0$. We are interested in estimating the magnitude of the interaction potential, as well as its   dependence on the momentum of the bound state when the magnon momentum is close to $k_0$.  
 
The scattering amplitude $V_{b\text{-}m}(k,p,q)$ is defined in Eq. (\ref{Vbm}). The  states with one magnon and one bound magnon pair  have the form 
\begin{equation}
 \ket{b,p;k} = \frac{1}{L}\sum_{j,l;r>0} e^{ikj} e^{ip(l+r/2)}\Psi(p,r)S^{-}_jS^{-}_{l}S^{-}_{l+r} \ket{\Uparrow}.
\end{equation}
We consider the   scattering problem on the infinite lattice, which  allows an analytic calculation of $V_{b\text{-}m}(p,k,q)$ in terms of  $\Psi(r)$.

\subsection{Setup}
\label{sec:orgec4d936}

We introduce a basis in the $N$-magnon subspace with total momentum $P$
denoted
\begin{multline}\label{Prstates}
 \ket{P; r_1,\ldots, r_{N-1}} =\\
= \sum_le^{iP[l + (N-1)r_1/N + (N-2)r_2/N + 
 \ldots + r_{N-1}/N] }\\
\times S^-_l S^-_{l+r_1}S^-_{l+r_1+r_2}\ldots S^-_{l+r_1+\ldots+r_{N-1}} \ket{\Uparrow},
\end{multline}
with \(r_i >0\) for all $i=1,\dots,N-1$. We define \(\ket{P,r_1,\ldots,r_{N-1}} = 0\)  when any \(r_i < 1\).  For instance,  in this notation  the bound-magnon-pair state ($N=2$) with momentum $p$ is expressed as $\ket{b,p} =\frac{1}{\sqrt{L}} \sum_{r>0}\Psi(p,r)\ket{p;r}$.  It is also convenient to  consider the transformation of the state in Eq. (\ref{Prstates}) under site inversion, $\mc {I}:\mathbf S_j\mapsto   \mathbf S_{-j}$. One can verify that \be
\mc I: \ket{P; r_1,\ldots, r_{N-1}} \mapsto  \ket{-P; r_{N-1},\ldots, r_{1}}. 
\ee
Taking the composition with complex conjugation, $\mc K:i\mapsto -i$, we obtain\be
\mc K  \mc I: \ket{P; r_1,\ldots, r_{N-1}} \mapsto  \ket{P; r_{N-1},\ldots, r_{1}}. 
\ee
For an arbitrary  state $|\psi\rangle$, we shall refer to the state  $\mc K  \mc I |\psi\rangle$ as the parity conjugate (p.c.) of  $|\psi\rangle$.

 Let us write the Hamiltonian
\begin{equation}
 H = J_1H_1 + J_2H_2,
\end{equation}
with \be
H_n= \sum_j\left[\frac{1}{2}(S^+_jS^-_{j+n} + S^-_jS^+_{j+n}) + S^z_{j}S^z_{j+n} - \frac14\right],
\ee
for $n=1,2$. In the calculation of effective scattering amplitudes, we omit the magnetic field term in Eq. (\ref{model}) because we work in a sector with fixed   $S^z_{\text{tot}}$. 
The action of the Hamiltonian is readily understood in the basis of Eq. (\ref{Prstates}). \edit{The terms that hop magnons  (stemming from the transverse part of the exchange interaction) lead to a phase determined by the change of the total center-of-mass momentum as well as a shift of the relative coordinates. Note that to hop a magnon to the right (left) one should increase (decrease) its relative coordinate with respect to the previous magnon  but one should also shift the subsequent coordinate in the opposite direction. The interaction terms stemming from the longitudinal part $S_j^zS_{j+n}^z$ simply count the number of magnons   separated by   $n$ sites. For the \(J_1\) terms, the hopping is straight forwardly implemented by $r_i\to r_i+1$ and $r_{i+1} \to r_{i+1}-1$  and the interaction is simply counting the number of \(r_i=1\). 
For the \(J_2\) terms, the basic processes are similar but with steps of two. However,  there are now two complications to take note of: first, in considering the  hopping term we have the possibility that magnons hop over each other. This leads to a hopping process when \(r_i=1\) and \(r_{i-1}>1\) that switches the relative coordinates \(r_i\) and \(r_{i-1}\). The net effect   is to shift $r_{i-1}\to r_{i-1}+ 1$,$\  r_{i+1}\to r_{i+1}-1$ for hopping  to the right over the   magnon  at  $l+\sum_{l\leq i}r_l$, and the opposite for hopping to the left. Secondly, for the interaction term one also needs to account separately for the case  when three magnons  are all adjacent, i.e. \(r_i=r_{i+1}=1\), since    then the two outer magnons are two lattice spacings apart and hence interact via $J_2$. }

The computation of the matrix element  allows a significant simplification: The result will be of the form
\begin{equation}
V = J_1V_1 + J_2V_2.    
\end{equation}
Here we can compute the parts $V_{1,2}$ as if $J_{2,1}=0$ while  $J_{1,2}=1$ as long as we keep the bound-state wave function as a formal function in all the equations.

\subsection{The $V_1$ term}

Since the total momentum $P=p+k$ is a good quantum number, hereafter we omit the dependence on $P$ and adopt the shorthand notation $|P;r_1,\dots,r_{N-1}\rangle\equiv |r_1,\dots,r_{N-1}\rangle$. We also omit the momentum dependence of the bound state wave function  and write $\Psi(p,r)\equiv \Psi(r)$. We write the scattering state of a  magnon and a bound magnon pair as 
\be
 \ket{b,p;k} =\frac1L( \ket{\psi_1} + \ket{\psi_2} + \ket{\psi_3}),
\ee
where 
\be
 \ket{\psi_i} = \sum_{r_1,r_2>0} \ket{\psi_i(r_1,r_2)},
 \ee
with
\be
 \ket{\psi_i(r_1,r_2)} =  \psi_i(r_1,r_2)\ket{r_1,r_2}.
\ee
The index $i=1,2,3$ 
corresponds to configurations with the free magnon on the right, left, and in the middle respectively, given by the wave functions 
\begin{align}
\label{eq:6}
\psi_1(r_1,r_2) & = e^{i(2k-p)(r_1 +2r_2)/6 }\Psi(r_1),\\
\psi_2(r_1,r_2) & = e^{i(p-2k)(2r_1 + r_2)/6}\Psi(r_2) ,\\
\psi_3(r_1,r_2) & = e^{i(2k - p)(r_1 - r_2)/6}\Psi(r_1+r_2).
\end{align}
We compute the action of \(\tilde{H}_1 =H_1 - \varepsilon^{(1)}(k) - \mathcal{E}^{(1)}_b(p)\) where $\varepsilon^{(1)}(k) = \cos(k)-1$ is the term in the magnon dispersion with coefficient  $J_1$ [i.e., the function $\varepsilon(k)$ that we  would   obtain if we set $J_1=1$ and $J_2=h=0$] and \be
\label{eq:2}
  \mathcal{E}^{(1)}_b(p) =-1+ \cos(p/2)\frac{\Psi(2)}{\Psi(1)}
  \ee 
 is the analogous term coming from the bound state dispersion. The other pieces of the dispersion relations that will be included in the $V_2$ term are $\varepsilon^{(2)}(k)=\cos(2k)-1$ and $ \mathcal{E}^{(2)}_b(p)=\cos(p)[1 + \Psi(3)/\Psi(1)]-2$. The latter is such that $J_1\mathcal{E}^{(1)}_b(p)$ and $J_2\mathcal{E}^{(2)}_b(p)$ add up to $\mc E_b(p)$ (with $h=0$) provided that $\Psi(r)$ is the wave function that makes $|b,p\rangle$ an eigenstate of $H$ in the $N=2$ sector. This follows from  $\bra{p;1}H\ket{b,p} = \mathcal{E}_b(p)\Psi(1)$.
In the following, we will treat $\Psi(r)$ as an input, but note that  it depends nontrivially  on   $J_1$ and $J_2$.  For $r>1$, we have
\be
 \mathcal{E}^{(1)}_b(p)\Psi(r) = -2 + \cos\left(\frac{p}{2}\right)\left[\Psi(r+1) + \Psi(r-1)\right].
\ee
The result after subtraction of the dispersion related terms is readily understood if we keep the picture of the free magnon and the bound state as two distinct particles in mind: the terms that survive are either due to the interaction  (longitudinal part of the Heisenberg exchange) between the free magnon and one of the magnons in the bound state  or correspond to hops obstructed by the presence of the other particle (the latter appearing with a minus sign).
This gives us
\begin{align}
\label{eq:3}
 \tilde{H}_1\ket{\psi_1} &= \sum_{r>0}\left[1\! -\! \frac{e^{ip/2}}{2} \frac{\Psi(r+1)}{\Psi(r)}\! -\! \frac{e^{-ik}}{2} \right]\ket{\psi_1(r,1)},\\
  \tilde{H}_1\ket{\psi_3} &=  \sum_{r>0}\left[1\! -\!\frac{e^{-ip/2}}{2} \frac{\Psi(r)}{\Psi(r+1)}\! -\! \frac{e^{ik}}{2} \right]\ket{\psi_3(r,1)}\nonumber
                  \\
              & \qquad    +
                  \mathrm{p.c.},
\end{align}
where p.c. denotes the parity conjugate.   The  terms from $\ket{\psi_2}$  are given by $\tilde{H}_1\ket{\psi_2}=\mc K\mc I\left( \tilde{H}_1\ket{\psi_1}\right)$. From these expressions the computation of  $V_1$ is tedious but  straightforward. We obtain
\begin{align}
\label{V1}
V_1 =& 8\sin\left(\frac{k}2\right)\sin\left(\frac{k}2\right)\sum_{r=2}^{\infty}\Psi(r)\Psi(r)\notag\\
&+ 8\sin\left(\frac{k}2\right)\sin\left(\frac{ p-k}2\right)\times\notag\\
&\times\sum_{r=1}^{\infty}\cos\left[\left(\frac{p}2 -k\right)r\right]\Psi(1)\Psi(r)  \notag\\
&+8\sin\left(\frac{k}2\right)\sin\left(\frac{p-k}2\right)\sum_{r=1}^{\infty}\Psi\left(r \right)\Psi\left(r+1\right).
\end{align}

\subsection{The \(V_2\) term}
\label{sec:org26cd2c1}
Following the same lines as for $V_1$, we compute the action of $\tilde{H}_2$ on the $\ket{b,p;k}$ state of a bound magnon pair and a free magnon. As in the $V_1    $ case, a physical picture allows to write down the result immediately. The states $\ket{\psi_{1,2,3}}$ correspond to the cases where the free magnon is on the right, on the left, or in the middle of the bound magnon pair. Thus, interpreting these states as a configuration of two particles in this way and acting with the Hamiltonian, we   find that the presence of the other particle can obstruct some possible moves   or add additional interactions in comparison with a situation in which the other particle would not be present. For instance, if the free magnon particle is two sites away from one of the magnons in the bound magnon pair, as is the case in e.g. $\ket{\psi_1(r,2)}$, the magnon cannot hop to the left and the right most magnon in the bound state cannot hop to the right. Furthermore, there is now an interaction between  the two particles. This leads to a term $[1 - \frac{1}{2}e^{-i2k} - \frac{1}{2}e^{ip}\Psi(4)/\Psi(2)]\ket{\psi_1(r,2)}$ in $\tilde{H}_2\ket{\psi_1}$. The first term on the righthand side stems from the interaction between the free magnon and one magnon in the bound magnon pair. The momentum dependent phases in the second and third terms can be identified from the obstructed \edit{hopping processes}: $e^{-i2k}$ corresponds to the hopping of the magnon two sites to the left, while $e^{ip}$ corresponds to the bound-state center of mass hopping one site to the right if one of its constituent magnons hops two sites to the right.  The ratio $\Psi(4)/\Psi(2)$ corresponds to   changing the separation between the magnons in the bound magnon pair from two to four sites.  In this way, all terms in $\tilde{H}_2\ket{\psi_i}$ are straightforward to write down.   We then compute the appropriate inner products to obtain the result  
\begin{align}
V_2 =&\ [8\sin(k)^2 + 4\sin(k)\sin(p-k)]\Psi(1)\Psi(1)\notag\\
&-\cos(5p/2-3k)\Psi(1)\Psi(2) +\cos(p)\Psi(1)\Psi(3)  \notag\\
&+\cos(p/2-3k)\Psi(2)\Psi(3)\notag\\
&+8\sin(k)\sin(k)\sum_{r=3}^{\infty}\Psi(r)\Psi(r) \notag\\
&+8\sin(k)\sin(p-k)\sum_{r=1}^{\infty}\Psi(r)\Psi(r+2) \notag\\
&+8\sin(k)\sin(p-k)\sum_{r=1}^{\infty}\cos\left[\left(\frac{p}2-k\right)r\right]\Psi(2)\Psi(r).  
\end{align}
Note that this reduces to the result for $V_1$ in Eq.~(\ref{V1}) if we put $(r,k,p) \to (2r,k/2,p/2)$ and declare $\Psi(r) = 0$ for odd $r$. This corresponds to the decoupling of the  $J_1$-$J_2$ spin chain upon putting $J_1 = 0$ in terms of two independent Heisenberg chains with doubled lattice spacing living on the even and odd sublattices. 

\subsection{The $V_{b\text{-}m}$ result}
For the final result, we evaluate
\begin{equation}
 V_{b\text{-}m}(p,k,0) = J_1V_1(p,k) + J_2V_2(p,k).    
\end{equation}
The bound-state wave function $\Psi(r)$ can be obtained numerically by solving the Hamiltonian in the $N=2$ subspace of momentum $p$. We set $k=k_0$ to compute the interaction with the free magnon at the minimum of the magnon dispersion. The result is shown in Fig. \ref{fig:V}. We note that, as mentioned in the main text, the precise value is rather sensitive to the details of the wave function $\Psi(r)$. Thus, we expect \edit{the} interaction in the effective model  of Eq. (\ref{bmeffectiveinter}) to be strongly renormalized for a finite density of bound magnon pairs. 

\section{Exponents from the mobile impurity model}
\label{app:exponents}
We outline the calculation of the exponents from the mobile impurity model for the reader's convenience. We refer to Ref.~\onlinecite{Imambekov2012} and references therein for  further details.

The starting point is the mobile impurity model Eq.~(\ref{Himp}) and an expression of a space and time dependent correlator such as
\be
\label{correlator}
C(x,t)= \langle e^{  i \sqrt{\pi} \phi(x,t)} a_F(x,t)a_F^{\dag}(0,0)e^{- i \sqrt{\pi} \phi(0,0)}\rangle. 
\ee
There are two important steps  in evaluating this expression: decoupling the impurity mode (magnon) from the low-energy modes and rescaling the bosonic fields (equivalent to a Bogoliubov transformation diagonalizing the TL liquid Hamiltonian). We start by decoupling the impurity by the unitary transformation 
\be
   U = \exp \left[ i \sqrt{\pi}\int dx (\kappa_1 \theta + \kappa_2 \phi)a_F^{\dag}a_F \right].
\ee
For any field $f$, we define the transformed field $\bar{f}=U^\dag fU$. 
This gives the following relations
\bea
  \partial_x\phi &=&  \partial_x\bar{\phi} - \sqrt{\pi} \kappa_1 a_F^{\dag}a_F\notag,\\
    \partial_x\theta &=&  \partial_x\bar{\theta} - \sqrt{\pi} \kappa_2 a_F^{\dag}a_F,\\
    a_F &=& \bar{a}_F e^{-i \sqrt{\pi}(\kappa_1\theta + \kappa_2\phi)}.\notag
\eea
Choosing 
\be
 \kappa_{1} = \frac{\gamma_1 K }{\pi},\qquad  \kappa_{2} = \frac{\gamma_2  }{\pi  K} ,   
\ee
we see that the \edit{interaction term cancels}. 
Additional terms which are generated are less relevant (the impurity model only contains marginal terms and the magnon mass term) or correspond to magnon-magnon interactions neglected because we only consider configurations with a single magnon. On the level of the correlator, we see that \edit{the unitary} decoupling means we have to attach a vertex operator of the bosonic modes to each  magnon operator.  
\edit{The expression} for the correlator (\ref{correlator}) factorizes in terms of the propagator $G(x,t)$ of a free particle, see Eq. (\ref{propagator}), multiplying a correlator expressed only in terms of the bosonic fields. This correlator can be evaluated in the standard way within TL liquid theory. For the example in Eq. (\ref{correlator}), we find\be
C(x,t)=G(x,t) [i(vt - x) +0^+]^{-\mu_R} [i(vt + x) +0^+]^{-\mu_L},
\ee
where
\be
 \mu_{R,L} = \left(\frac{\sqrt{K}}{2 } - \frac{\gamma_2}{2\pi \sqrt{K}} \pm \frac{\gamma_1 \sqrt{K}}{2\pi}\right)^2.
\ee
Taking the Fourier transform leads to the expression of the   threshold exponent  in the frequency domain, $S(q,\omega)\sim (\omega-vq)^{\mu}$, with 
\be
\mu = \mu_R + \mu_L - 1.
\ee
This example gives the expression for $\mu^{+-}_+$, see Eq. (\ref{mu+-}). In a similar calculation for $\mu^{+-}_-$, we find the result with $\gamma_2 \to - \gamma_2$ responsible for the asymmetry between the minima at $q = k_0 \pm \pi \rho_0$ in $S^{+-}(q,\omega)$. The calculation of the exponents $\mu^{-+}_{\pm}$ associated with the threshold $S^{-+}(q,\omega)$ is similar but in this case one must use the representation  of the spin operator in   Eq. \eqref{eq:projSp}.
\end{document}